\documentclass[10pt,doublecolumn,twoside,journal]{IEEEtran}
\usepackage{graphicx}
\usepackage{algorithm}
\usepackage{algorithmicx}  
\usepackage{algpseudocode}

\usepackage{epsfig,latexsym}
\usepackage{float}
\usepackage{indentfirst}
\usepackage{amsmath}
\allowdisplaybreaks[4]
\usepackage{amssymb}
\usepackage{times}
\usepackage{subfigure}
\usepackage{pifont}
\usepackage{psfrag}
\usepackage{cite}
\usepackage{url}
\usepackage{lastpage}
\usepackage{bm}
\usepackage{color}
\usepackage{fancyhdr}
\usepackage{caption}
\usepackage{balance}

\newtheorem{Proposition}{Proposition}

\newtheorem{Lemma}{Lemma}

\newtheorem{Remark}{Remark}

\ifCLASSINFOpdf  
\else
\fi

\usepackage[utf8]{inputenc}
\begin{document}
	\title{Exploiting Deep Learning for Secure Transmission in an Underlay Cognitive Radio Network}
\author{Miao Zhang, \IEEEmembership{Member, IEEE,} Kanapathippillai Cumanan, \IEEEmembership{Senior Member, IEEE,}  Jeyarajan Thiyagalingam, \IEEEmembership{Senior Member, IEEE,} Yanqun Tang, Wei Wang, \IEEEmembership{Member, IEEE,} Zhiguo Ding, \IEEEmembership{Fellow, IEEE,} and Octavia A. Dobre, \IEEEmembership{Fellow,~IEEE}
	\thanks{The work of M. Zhang was supported by the Research Start Up Funding of Chongqing Jiaotong University under grant number 2020020070. The work of Y. Tang was supported by the Guangdong Natural Science Foundation under grant number 2019A1515011622 and the National Natural Science Foundation under grant number 62071499. The work of W. Wang was supported in part by the Six Categories Talent Peak of Jiangsu Province under Grant KTHY-039, the Science and Technology Program of Nantong under Grant MS22019019 and the Verification Platform of Multi-tier Coverage Communication Network for oceans under Grant LZC0020. ({\it Corresponding author: Yanqun Tang.})}
	\thanks{M. Zhang is with the School of Information Science and Engineering, Chongqing Jiaotong University, Chongqing, China, (email: miao.zhang@cqjtu.edu.cn).}
	\thanks{K. Cumanan is with the Department of Electronic Engineering, University of York, York, United Kingdom, YO10 5DD (email:  kanapathippillai.cumanan@york.ac.uk).}
	\thanks{J. Thiyagalingam is with the Scientific Computing Department of Rutherford Appleton Laboratory, Science and Technology	Facilities Council, Harwell Campus, Ditcot, UK (email: t.jeyan@stfc.ac.uk).}
	\thanks{Y. Tang is with the School of Information Science and Engineering, Chongqing Jiaotong University, Chongqing, China, also with the School of Electronics and Communication Engineering, Sun Yat-Sen University, Shenzhen, 510006, China (email: tangyq8@mail.sysu.edu.cn)}
	\thanks{W. Wang is with the School of Information Science and Technology, Nantong University, Nantong, China, and with the Nantong Research Institute for Advanced Communication Technologies, Nantong, China, and also with Research Center of Networks and Communications, Peng Cheng Laboratory, Shenzhen, China (e-mail: wwang2011@ntu.edu.cn).}
	\thanks{Z. Ding is with the School of Electrical and Electronic Engineering, The University of Manchester, Manchester, UK (email: zhiguo.ding@manchester.ac.uk).}
	\thanks{O. A. Dobre is with the Department of Electrical and Computer Engineering, Memorial University, St. John’s, NL A1B 3X5, Canada (email: odobre@mun.ca).}
}
	\maketitle

\begin{abstract}
This paper investigates a machine learning-based power allocation
design for secure transmission in a cognitive radio (CR) network. In
particular, a neural network (NN)-based approach is proposed to
maximize the secrecy rate of the secondary receiver under the
constraints of total transmit power of secondary transmitter,
and the interference leakage to the primary receiver, within which three different regularization schemes are developed. The key advantage of the proposed algorithm over conventional approaches is the capability to solve the power allocation problem with both perfect and
imperfect channel state information. In a conventional setting,
two completely different optimization frameworks have to be designed,
namely the robust and non-robust designs. Furthermore, conventional
algorithms are often based on iterative techniques, and hence, they
require a considerable number of iterations, rendering them less
suitable in future wireless networks where there are very stringent
delay constraints. To meet the unprecedented requirements of future ultra-reliable
low-latency networks, we propose an NN-based approach that can
determine the power allocation in a CR network with significantly
reduced computational time and complexity. As this trained NN only
requires a small number of linear operations to yield the required
power allocations, the approach can also be extended to different delay sensitive applications and services in future wireless
networks. When evaluate the proposed method versus conventional approaches, using a suitable test set, the proposed approach can achieve more than 94\% of the secrecy rate performance with less than 1\% computation time and more than 93\% satisfaction of interference leakage constraints. These results are obtained with significant reduction in computational time, which we believe that it is suitable for future real-time wireless applications.

\end{abstract}

	%
	\IEEEpeerreviewmaketitle
	
\section{Introduction}
Wireless communications have become an indispensable part of daily
life of people as they play a crucial role in our day-to-day activities
and the means of interactions in the current networked
society. However, information security is one of the major challenges
in wireless networks due to the open nature of wireless signal
transmission which is more vulnerable for interception and
eavesdropping. The conventional security methods employed at upper
layers in the current communication systems completely rely on
cryptographic techniques \cite{cumanan2014secrecy,chu2015robust}. Despite the fact that existing conventional security
techniques, developed based on some high complex intractable
mathematical problems, are difficult to break or intercept, the
broadcast nature of wireless transmissions introduces different
challenges in terms of secret key exchange and distributions
\cite{chu2016secrecy, zhang2017secure}. As a result, information
theoretic based physical layer security has been proposed to
complement the conventional cryptographic methods and to provide
additional security measures in wireless transmissions. Furthermore,
this approach exploits the dynamics of physical layer characteristics
of wireless channels to establish secure transmission
\cite{cumanan2014secrecy}. A reasonable secrecy rate can be realized
through physical layer security technique provided that the signal-to-interference plus noise ratio (SINR) of the channel of the legitimate
user is better than that of the channel of the eavesdropper
\cite{Shannon}. This novel technique was first theoretically proved by
Shannon \cite{Shannon} and then secrecy capacities of wiretap and
related channels were developed by Wyner \cite{Wyner} and Csiszar
\cite{csiszar}. In contrast to the conventional cryptographic methods,
physical layer security schemes are more suitable for practical
implementations as these techniques do not require any secret key
distributions or exchange. Furthermore, it is difficult for
interceptors to decipher the information transmitted across wireless
channels based on physical layer security \cite{chu2016secrecy}.

Recently, machine learning techniques have been applied widely as a solution approach to solve different challenging problems that have complicated structures with stringent constraints on computational time \cite{jiang2016machine}. Furthermore, artificial intelligence has become one of the fastest growing techniques in many research topics \cite{goodfellow2016deep} and its practical implementations can be realized through different machine learning techniques. As such, these techniques enable machines to acquire knowledge from their computations and make decisions according to the environment \cite{andrieu2003introduction,sebastiani2002machine,bishop2006}. There are various machine learning frameworks available in the literature \cite{james1995, goodfellow2016deep,andrieu2003introduction,sebastiani2002machine,bishop2006}, such as linear regression, logistic regression, and neural network (NN). NN is one of the well-known machine learning technique due to its capabilities to simply realize different relationships in complicated and statistical data sets \cite{chen2017machine,demuth2014neural}. In recent years, numerous research interests have been developed to utilize NN to design and optimize wireless communication systems, where the researchers believe that NN will be the core technique for 5G and beyond wireless systems \cite{8847377,8618345,8782877,8761962}.


	\subsection{Motivation and Contributions}
	In secure transmission designs, different optimization approaches with various approximations techniques have been widely exploited to solve complicated and mathematically intractable resource allocation problems \cite{zhang2018robust, cumanan2017physical, cumanan2016secrecy,chu2016simultaneous,chu2017robust,8649584}. However, these techniques often have been developed based on iterative approaches to yield either optimal or sub-optimal solutions. The computational complexities associated with these conventional optimization techniques are neither affordable in low powered devices in Internet-of-Things (IoT) nor suitable for applications with ultra reliability and low latency in future wireless networks. Furthermore, these conventional optimization techniques pose different challenges in delay sensitive systems as the dynamic nature of real-time parameters requires frequent updates in very short time \cite{sun2017learning}. This introduces different stringent delay requirements in updating those design parameters which is impossible to meet by conventional optimization approaches. Machine learning techniques can be considered as the potential solution approaches to solve these real-time update issues. Among a number of machine learning approaches, the deep learning approach has a number of benefits. Although some of these benefits are shared across different methods, deep learning offers better learn-ability with increased volumes of data.  We summarize these benefits as follows: 
		
		\begin{enumerate}
			\item NN has the potential capabilities to provide a solution with a short time frame with reduced computational complexity \cite{zhang2019deep}, compared to other  machine learning techniques, such as support vector machine (SVM) and Gaussian processes (GP)~\cite{domingos2012few,zhang2019deep}. A particular advantage here is that conventional machine learning  approaches use all available data, whereas the NN relies on samples of data from batches (mini-batch gradient descent algorithm). This process demands only a sub-set of the available large dataset at each training step, opposed to every data point;
			
			\item A single NN model can be trained to meet the objectives of multiple tasks~\cite{zhang2019deep}, whereas it is difficult for other machine learning techniques to achieve those multi-objectives with the same model; and
			
			\item Furthermore, NN is able to automatically extract features from the data with highly complex datasets and to formulate the latent representations, which can further help with learning~\cite{lecun2015deep}.
		\end{enumerate}

	In the literature, several bodies of work have demonstrated that machine learning techniques can be exploited to solve these types of problems in different real-time wireless communication applications. For example, deep learning-based channel estimation and signal detection techniques in orthogonal frequency division multiplexing (OFDM) systems is investigated in \cite{ye2018power}. A deep NN-based method for efficient on-line configuration of reconfigurable intelligent surfaces is proposed in \cite{huang2019indoor}, where the transmitted signal focusing is improved under the indoor environment. The deep reinforcement learning based joint transmit beamforming and phase shift matrix design for reconfigurable intelligent surface aided MISO systems is studied in \cite{huang2020reconfigurable}. The NN-based spectrum and energy efficiency maximization techniques is proposed for cognitive radio (CR) network in \cite{zhou2018resource}. A learning-based approach for wireless resource management is presented in \cite{sun2017learning}, whereas a reinforcement learning based resource allocation technique is developed for vehicle-to-vehicle communications in \cite{ye2018deep}. A deep NN is utilized to learn the interference management over interference-limited channels in \cite{sun2018learning}, whereas the authors design a deep NN for channel calibration between the uplink and downlink directions in generic massive MIMO systems in \cite{huang2019deep}. However, none of these work have considered employing machine learning techniques to simultaneously solve resource allocation problems with perfect and imperfect channel state information (CSI) in secure communication systems.
	
    In general, the motivations behind this work can be summarized as follows: (1) Although the conventional optimization approaches can yield global or local optimal solutions for resource allocation problems, the nature of their complex implementations render them less practical for real-world deployments, particularly on resource-limited edge devices where tolerance for delays are very minimal.  
	 	For NNs, once trained, the inference step does not demand back-propagation, at which point it only relies on limited number of floating point arithmetic. This offers a two-fold benefit. First, once trained (using powerful computational resources), the NN model can be moved around for inference purposes, particularly on edge devices where computational resources are often limited. Secondly, an NN-based approach  can offer almost near real-time performance. This is clearly evidenced by modern edge devices, such as smart phones and cameras. (2) As NN provides reduced computational complexity for inference compared to other machine learning  techniques, it is useful for our resource allocation problem. (3) Finally, with the rise of machine learning  algorithms in various domains of sciences, the community would benefit if some baseline performance can be established for resource allocation problems in wireless communications and be compared against conventional optimization-based solutions. 
	 	
	 	To carry out the study, in this paper, we consider a secure transmission in a CR network problem as shown in Fig. 1. This secure network consists of one primary transmitter (PU-Tx), one primary receiver (PU-Rx), one secondary transmitter (SU-Tx), one secondary receiver (SU-Rx) and one eavesdropper (EVE). These terminals are equipped with a single antenna. Our main objective is to design an NN approach that can achieve near optimal secrecy rate performances with significantly reduced computational time compared to the existing conventional optimization schemes in the literature. In particular, the optimal power allocation is determined to maximize the achievable secrecy rate under the constraints of total transmit power of the SU-Tx and the interference leakage to the PU-Rx. We develop two approaches in this paper: the conventional optimization approach and NN-based framework. We show that the NN-based approach can be exploited to solve both robust and non-robust secrecy rate maximization problem, whereas the conventional optimization techniques require completely two different problem formulation and solution approaches. Our contributions of this work are summarized as follows:
	 	\begin{enumerate}	
	 		\item Firstly, to the best of our knowledge and surveys~\cite{zhang2019deep,sun2020machine,al2020survey}, none of the existing work considered developing an NN framework to solve the secrecy rate maximization problems in an underlay CR network. 
	 		
	 		\item Secondly, due to the imperfections and non-linearities in practical systems \cite{o2017introduction}, we also consider a more practical imperfect CSI scenario in this paper, whereas most of the previous works that apply NN for resource allocation problems only consider the perfect CSI scenarios. Therefore, the framework of the proposed NN is  different from those found in related works, i.e., we have added the channel error bounds as input parameters to enable the NN to learn the impact of these errors on the power allocation.
	 		
	 		\item Thirdly, we propose an NN-based algorithm to simultaneously solve the secrecy rate maximization problem with perfect and imperfect CSI at the SU-Tx. The key advantage of the developed approach is that the same NN-based algorithm can be exploited to solve both the robust and the non-robust secrecy rate maximization problems with the both imperfect and perfect CSI, respectively. Opposite to that, in the conventional optimization approaches, these problems need to be formulated into two completely different optimization frameworks. Furthermore, to reduce over-fitting, we also embed two regularization techniques into our proposed NN designs. To generate the required training set, we utilize the conventional optimization framework and then train the NN with this training set to determine appropriate weights of the connections in the proposed NN. These weighted connections establish a mathematical relationship between the input and the corresponding output. After completing the training process, we evaluate the performance of the proposed NN-based approach versus the the conventional optimization approaches available in the literature. 
	 		
	 		\item Finally, we compare the performance of both schemes in terms of achieved secrecy rate and required computation time to demonstrate the effectiveness and superiority of our proposed NN scheme.
	 		
 	\end{enumerate}

	The remainder of this paper is organized as follows. The system model is presented in Section II, whereas the secrecy rate maximization problems with both perfect and imperfect CSI are formulated and solved by using conventional optimization technique in Section III. Section IV presents an NN-based optimization framework. Section V provides simulation results to demonstrate the effectiveness of the proposed approach. Section VI discusses the limitations of the proposed approach and several potential directions for future work, and finally, Section VII concludes this paper.
	\subsection{Notations}
	We use the upper and the lower case boldface letters for matrices and vectors, respectively. $(\cdot)^{-1}$, $(\cdot)^T$ and $(\cdot)^H$ stand for inverse, transpose and conjugate transpose operation, respectively. $|a|$ represents the absolute value of $a$. $[x]^{+}$ defines $\max\{x, 0\}$. The 1-norm and 2-norm of $x$ are expressed respectively as $||x||_{1}$ and $||x||_{2}$. $\mathbf{A}\cdot\mathbf{B}$ represents the dot product of matrix $\mathbf{A}$ and $\mathbf{B}$. $h^{'}(x)$ is the first derivative of function $h$ at $x$. The circularly symmetric complex Gaussian distribution with mean $\mu$ and variance $\sigma^{2}$ is represented by $\mathcal{CN}(\mu,\sigma^2)$.
	\section{System Model and Problem Formulation}
	\begin{figure}[t]
		\centering\includegraphics[width=\linewidth]{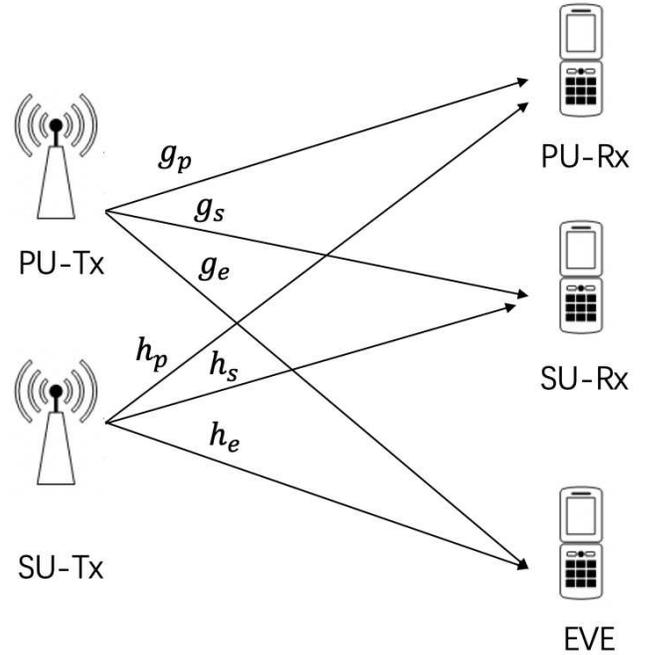}
		\caption{A CR network with one PU-Tx, one PU-Rx, one SU-Tx, one SU-Rx and one EVE. Each is equipped with a single antenna.}
	\end{figure}

	We consider a CR network as shown in Fig. 1 with five terminals: one PU-Tx, one SU-Tx, one SU-Rx, one PU-Rx and one EVE. All terminals are equipped with single antenna. The SU-Tx intends to send a confidential message to the SU-Rx while ensuring that the interference leakage to the PU-Rx is less than a predefined threshold. At the same time, the EVE attempts intercepting the information sent by the SU-Tx to SU-Rx. The channels between PU-Tx and PU-Rx, SU-Rx, and EVE are represented by $g_{p}$, $g_{s}$, and $g_{e}$, respectively, whereas the channels between the SU-Tx and PU-Rx, SU-Rx, and EVE are denoted by $h_{p}$, $h_{s}$, and $h_{e}$, respectively. The received signal at the SU-Rx and EVE can be expressed, respectively, as
	\begin{align}
     y_{s}=\sqrt{P_{s}}h_{s}x_{s}+\sqrt{P_{p}}g_{s}x_{p}+n_{s},
	\end{align}
	\begin{align}
	y_{e}=\sqrt{P_{s}}h_{e}x_{s}+\sqrt{P_{p}}g_{e}x_{p}+n_{e},
	\end{align}
    where $x_{s}(\mathbb{E}\{|x_{s}|^{2}\}=1)$ and $x_{p}(\mathbb{E}\{|x_{p}|^{2}\}=1)$ are the symbols sent from the SU-Tx to SU-Rx and the PU-Tx to PU-Rx, respectively. The noise at the SU-Rx and EVE are denoted by $n_{s}(\mathbb{E}\{|n_{s}|^{2}\}=\sigma_{s}^{2})$ and $n_{e}(\mathbb{E}\{|n_{e}|^{2}\}=\sigma_{e}^{2})$, respectively. Furthermore, $P_{s}$ and $P_{p}$ represent the power allocations at the SU-Rx and the EVE, respectively. The SINR at the SU-Rx and EVE are defined as
    \begin{align}
    \gamma_{s}=\frac{P_{s}|h_{s}|^{2}}{P_{p}|g_{s}|^{2}+\sigma_{s}^{2}},
    \end{align}
    \begin{align}
    \gamma_{e}=\frac{P_{s}|h_{e}|^{2}}{P_{p}|g_{e}|^{2}+\sigma_{e}^{2}}.
    \end{align}
    The achievable secrecy rate at the SU-Rx can be written as \cite{zhang2016secrecy}
    \begin{align}
    R_{s}=[\log_{2}(1+\gamma_{s})-\log_{2}(1+\gamma_{e})]^{+}.
    \end{align}
    The interference leakage to the PU-Rx can be expressed as
    \begin{align}
    P_{in}=P_{s}|h_{p}|^{2}.
    \end{align}
With these definitions, the secrecy rate maximization problem can be formulated as
\begin{align}\label{Sec_Ori}
	\max_{P_{s}}~~& R_{s}\nonumber\\
	s.t.~~ &P_{s}|h_{p}|^{2}\leq q,\nonumber\\
	~~&P_{s}\leq P_{t}, P_{s}\geq 0,
\end{align}
where $q$ is the maximum interference leakage to the PU-Rx, and $P_{t}$ is the maximum transmit power available at the SU-Tx. In the following sections, we present two ways to solve this problem: conventional optimization approaches and NN-based approach.
    \section{Conventional Optimization based Power Allocation Approach}
    In this section, we present conventional convex optimization approaches to solve the secrecy rate maximization problem defined in (\ref{Sec_Ori}) by taking into account the scenarios of having both perfect and imperfect CSI at the SU-Tx.
    \subsection{Perfect CSI}
	In this subsection, we present the conventional convex optimization-based approach to solve the problem defined in (\ref{Sec_Ori}) with perfect CSI assumption. The original problem (\ref{Sec_Ori}) is non-convex in its original form due to the non-convex objective function. Based on the monotonicity of logarithmic functions, we reformulate the original problem in (\ref{Sec_Ori}) as
	\begin{align}
	\max_{P_{s}}~~&\frac{1+\frac{P_{s}|h_{s}|^{2}}{P_{p}|g_{s}|^{2}+\sigma_{s}^{2}}}{1+\frac{P_{s}|h_{e}|^{2}}{P_{p}|g_{e}|^{2}+\sigma_{e}^{2}}}\nonumber\\
	s.t.~~&P_{s}|h_{p}|^{2}\leq q,\nonumber\\
	~~&P_{s}\leq P_{t}, P_{s}\geq 0.
	\end{align}
	The above problem still remains non-convex due to the fractional objective function, and therefore, it cannot be directly solved using existing convex optimization tools. To circumvent this non-convexity issue, we convert the original problem into a two-level optimization problem, namely outer problem and inner problem. The outer problem can be written with respect to (w.r.t.) a new scalar variable $t$ as
	\begin{align}\label{Sec_Outer}
	\max_{t\geq 0}&~~\frac{f(t)}{1+t},
	\end{align}
	whereas the inner problem can be expressed for a given $t$ as
	\begin{align}\label{Sec_inner}
	f(t)=\max_{P_{s}}~~&1+\frac{P_{s}|h_{s}|^{2}}{P_{p}|g_{s}|^{2}+\sigma_{s}^{2}}\nonumber\\
	s.t.~~ &P_{s}|h_{p}|^{2}\leq q,\nonumber\\
	~~&\frac{P_{s}|h_{e}|^{2}}{P_{p}|g_{e}|^{2}+\sigma_{e}^{2}}\leq t,\nonumber\\
	~~&P_{s}\leq P_{t}, P_{s}\geq 0.
	\end{align}
	The inner problem in (\ref{Sec_inner}) is convex for a given $t$ and can be solved by using standard interior-point methods. Since the inner problem in (\ref{Sec_inner}) is a convex problem, the outer problem in (\ref{Sec_Outer}) is a quasi-convex optimization problem w.r.t. variable $t$. Therefore, we employ a one-dimensional search to obtain the optimal $t^{*}$ and $P_{s}^{*}$\cite{cumanan2017secure}. The proposed one-dimensional search algorithm is summarized in Algorithm 1.
	\begin{algorithm}
		\caption{\!\!: One-dimensional search based on bisection method}\label{alg:golden}	\fontsize{10pt}{12pt}\selectfont
		\begin{algorithmic}[1]
			\State Initialize $t\in[0,t_{max}]$, $c=(\sqrt{5}-1)/2$, $a=0$, $b=t_{max}$, $t_{1}=(1-c)b$, $t_{2}=cb$;
			\State Compute $f(t_{1})$, $f(t_{2})$;
			\Repeat 
			\State  If~~ $\frac{1+f(t_{1})}{1+t_{1}}>\frac{1+f(t_{2})}{1+t_{2}}$, $b=t_{2}$, $t_{2}=t_{1}$, $f(t_{2})=f(t_{1})$, $t_{1}=a+c(b-a)$ and update $f(t_{1})$;
			\State  Else, ~$a=t_{1}$, $t_{1}=t_{2}$, $f(t_{1})=f(t_{2})$, $t_{2}=a+c(b-a)$, and update $f(t_{2})$;
			\Until $|b-a|\leq \epsilon$, where $\epsilon$ is threshold to terminate the algorithm.
		\end{algorithmic}
	\end{algorithm}
\subsection{Imperfect CSI}

		    In this subsection, we develop a tractable approach to solve the secrecy rate maximization problem with imperfect CSI available at the SU-Tx. We reformulate this robust problem into a tractable one by exploiting the Charnes-Cooper transformation \cite{charnes1962programming} and S-Procedure \cite{boyd1994linear}. 
			
			In practical scenarios, it is difficult for SU-Tx to obtain perfect CSI due to the channel estimation and quantization errors \cite{1199945}. Instead, the SU-Tx has knowledge of its estimated CSI and the uncertainty regions that contain the actual channel realizations, which is referred to imperfect CSI. Note that the imperfect CSI of the PU-Rx and SU-Rx can be estimated based on the standard CSI feedback techniques \cite{geraci2012secrecy}. Furthermore, the imperfect CSI of EVE can be obtained by the following two methods: (1) for the case when EVE is part of the system, the CSI can be estimated with the standard CSI feedback techniques, as is still part of the system and should be able to cooperate with SU-TX with its CSI feedback \cite{geraci2012secrecy}; (2) when EVE is not part of the system, the CSI can be estimated at the SU-Tx through the local oscillator power leakage from the EVE's RF front end, the details of which can be found in \cite{mukherjee2012detecting}.
	
		In this work, the imperfect CSI is modelled based on the deterministic models \cite{cumanan2014secrecy, 1199945, 1468482}, in which it is assumed that the actual channel lies in an ellipsoid centred at the channel mean. In this CSI assumption, the estimated CSI and the error bounds are known at the SU-Tx, while the actual value of channel errors are unknown \cite{cumanan2014secrecy, 1199945, 1468482}. The actual channel coefficients can be modelled with corresponding channel uncertainties as follows:
		\begin{align}
		h_{s}=\hat{h}_{s}+e_{s},~h_{e}=\hat{h}_{e}+e_{e},~h_{p}=\hat{h}_{p}+e_{p},
		\end{align}
		where $\hat{h}_{s}$, $\hat{h}_{e}$ and $\hat{h}_{p}$ are the channel coefficients estimated by the SU-Tx. Furthermore, the symbols $e_{s},~e_{e}$ and $e_{p}$ represent the channel uncertainties. These channel uncertainties are assumed to be bounded by a predefined ellipsoids, as follows:
		\begin{align}
		&|e_{s}|=|h_{s}-\hat{h}_{s}|\leq \epsilon_{s},\label{uncertain_hs}\\
		&|e_{e}|=|h_{e}-\hat{h}_{e}|\leq \epsilon_{e},\\
		&|e_{p}|=|h_{p}-\hat{h}_{p}|\leq \epsilon_{p},
		\end{align}
		where $\epsilon_{s}\geq0$, $\epsilon_{e}\geq0$ and $\epsilon_{p}\geq0$ are the error bounds. Based on these bounded channel uncertainties and the monotonicity of log functions, the robust secrecy rate maximization problem can be reformulated into the following robust optimization framework:
\begin{align}\label{robust_ori}
\max_{P_{s}}~~&\frac{1+\frac{P_{s}|\hat{h}_{s}+e_{s}|^{2}}{P_{p}|g_{s}|^{2}+\sigma_{s}^{2}}}{1+\frac{P_{s}|\hat{h}_{e}+e_{e}|^{2}}{P_{p}|g_{e}|^{2}+\sigma_{e}^{2}}}\nonumber\\
s.t.~~&P_{s}|\hat{h}_{p}+e_{p}|^{2}\leq q,\nonumber\\
~~&P_{s}\leq P_{t}, P_{s}\geq 0.
\end{align}
First, we introduce the Charnes-Cooper transformation  \cite{charnes1962programming} as
\begin{align}\label{CC_trans}
\overline{P}_{s}=\frac{P_{s}}{t},
\end{align}
 to recast the problem defined in (\ref{robust_ori}) as
 \begin{align}\label{robust_cc}
 \max_{\overline{P}_{s},~t}~~&t+\frac{\overline{P}_{s}|\hat{h}_{s}+e_{s}|^{2}}{P_{p}|g_{s}|^{2}+\sigma_{s}^{2}}\nonumber\\
 s.t.~~&t+\frac{\overline{P}_{s}|\hat{h}_{e}+e_{e}|^{2}}{P_{p}|g_{e}|^{2}+\sigma_{e}^{2}}\leq 1,\nonumber\\
 ~~&\overline{P}_{s}|\hat{h}_{p}+e_{p}|^{2}\leq tq,\nonumber\\
 ~~&\overline{P}_{s}\leq tP_{t}, \overline{P}_{s}\geq 0.
 \end{align}
The problem defined in (\ref{robust_cc}) can be rewritten by introducing a new slack variable $\tau$ and defining it in the epigraph form as
	\begin{subequations}\label{robust_epi}
		\begin{align}
		\max_{\overline{P}_{s},~t,~\tau}~~&\tau\\
		s.t.~~&t+\frac{\overline{P}_{s}|\hat{h}_{s}+e_{s}|^{2}}{P_{p}|g_{s}|^{2}+\sigma_{s}^{2}}\geq \tau,\label{robust_C1}\\
		~~&t+\frac{\overline{P}_{s}|\hat{h}_{e}+e_{e}|^{2}}{P_{p}|g_{e}|^{2}+\sigma_{e}^{2}}\leq 1,\label{robust_C2}\\
		~~&\overline{P}_{s}|\hat{h}_{p}+e_{p}|^{2}\leq tq,\label{robust_C3}\\
		~~&\overline{P}_{s}\leq tP_{t}, \overline{P}_{s}\geq 0.
		\end{align}	
	\end{subequations}
	The above problem is still intractable due to the infinite number of the uncertainty sets in the constraints (\ref{robust_C1})-(\ref{robust_C3}). To address this issue, we employ the following proposition:
	\begin{Proposition}:  \label{proposition:S_procedure}The constraints in  (\ref{robust_C1})-(\ref{robust_C3}) can be equivalently written as
		\begin{align}\label{SP_C1}
		&\left[\
		\begin{matrix}
		\lambda_{1}+\frac{\overline{P}_{s}}{P_{p}|g_{s}|^{2}+\sigma_{s}^{2}} & \frac{\overline{P}_{s}\hat{h}_{s}}{P_{p}|g_{s}|^{2}+\sigma_{s}^{2}}\\
		\frac{\overline{P}_{s}\hat{h}_{s}}{P_{p}|g_{s}|^{2}+\sigma_{s}^{2}} & \frac{\overline{P}_{s}|\hat{h}_{s}|^{2}}{P_{p}|g_{s}|^{2}+\sigma_{s}^{2}}+t-\tau-\lambda_{1}\epsilon_{s}^{2}
		\end{matrix}
		\right]\succeq \mathbf{0},\nonumber\\
		&\lambda_{1}\geq 0,
		\end{align}
		\begin{align}\label{SP_C2}
		&\left[\
		\begin{matrix}
		\lambda_{2}-\frac{\overline{P}_{s}}{P_{p}|g_{e}|^{2}+\sigma_{s}^{2}} & -\frac{\overline{P}_{s}\hat{h}_{e}}{P_{p}|g_{e}|^{2}+\sigma_{s}^{2}}\\
		-\frac{\overline{P}_{s}\hat{h}_{e}}{P_{p}|g_{e}|^{2}+\sigma_{s}^{2}} & 1-\frac{\overline{P}_{s}|\hat{h}_{s}|^{2}}{P_{p}|g_{e}|^{2}+\sigma_{e}^{2}}-t-\lambda_{2}\epsilon_{e}^{2}
		\end{matrix}
		\right]\succeq \mathbf{0},\nonumber\\
		&\lambda_{2}\geq 0,
		\end{align}
		and 
		\begin{align}\label{SP_C3}
		&\left[\
		\begin{matrix}
		\lambda_{3}-\overline{P}_{s} & -\overline{P}_{s}\hat{h}_{p}\\
		-\overline{P}_{s}\hat{h}_{p} &tq- \overline{P}_{s}|\hat{h}_{p}|^{2}+\sigma_{s}^{2}-\lambda_{e}\epsilon_{p}^{2}
		\end{matrix}
		\right]\succeq \mathbf{0},\nonumber\\
		&\lambda_{3}\geq 0.
		\end{align}
		\begin{IEEEproof}
			Please refer to Appendix A.
		\end{IEEEproof}
	\end{Proposition}

~\\

Therefore, we rewrite the problem in (\ref{robust_epi}) into the following equivalent form:
\begin{align}\label{robust_final}
\max_{\overline{P}_{s},~t,~\tau}~~&\tau\nonumber\\
s.t.~~&\textrm{(\ref{SP_C1})-(\ref{SP_C3})},\nonumber\\
~~&\overline{P}_{s}\leq tP_{t}, \overline{P}_{s}\geq 0.
\end{align}	
The above problem is convex, and therefore, the optimal $P_{s}^{*}$ can be obtained efficiently by the convex optimization tool box \cite{boyd2004convex}.

\section{Power Allocation Framework based on NN}
		\begin{figure*}[ht]
		\centering\includegraphics[width=5.4 in]{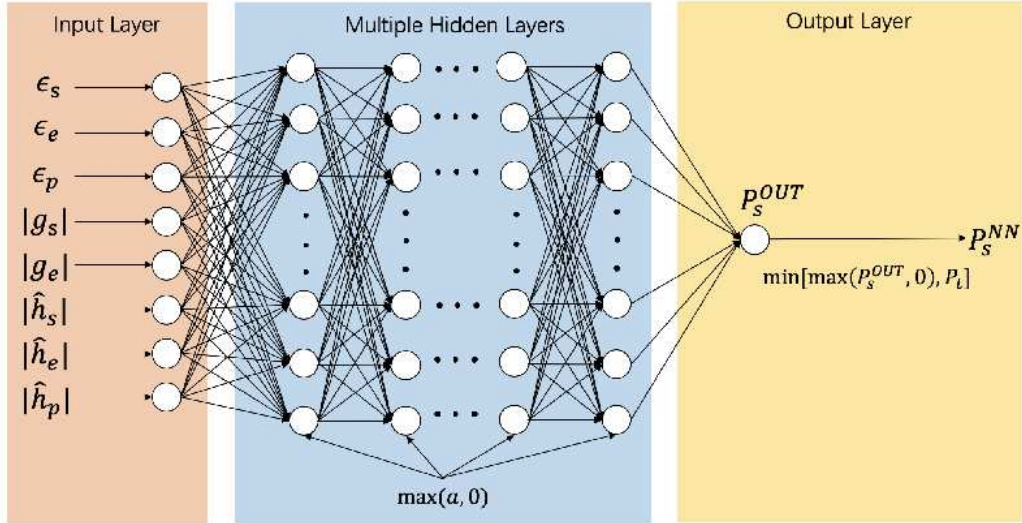}
		\caption{The structure of proposed NN.}	
		\label{fig:SRP}
\end{figure*}

In this section, we present our proposed NN-based schemes. In this approach, the secrecy rate maximization problem is treated as an unknown non-linear mapping, and an NN is trained to learn the relationship between the input and the output parameters.
		
		First, NNs can be considered as universal function approximators \cite{hornik1989multilayer}  and shown to have  remarkable capabilities of algorithmic learning \cite{reed2015neural}. As such, they are akin to conventional optimizer-based solutions. Second, the literature demonstrate that NN schemes have the capability to substantially reduce the computational complexity, and processing time for a variety of problems in wireless communications, such as, resource allocation \cite{zhou2018resource,sun2018learning,8626195}, channel estimation and signal detection \cite{ye2018power}, and physical layer designs \cite{o2017introduction}. Third, once the networks are trained (ideally using scalable computational resources), the resulting model is suitable for inference in very limited resource for real-time applications~\cite{jiang2017machine}.
	
		\begin{Remark} Although the notion of function approximation is useful to derive a powerful learned model, rendering an adaptive learning model is a challenging goal, including, but not limited to, anticipating varying inputs, noisy conditions, and failures.  As such, a simple NN-based approach alone cannot handle dynamic problems effectively. Instead, a solution to a dynamic problem will involve a hybrid approach, covering optimization techniques, NN, on-line learning, reinforcement learning and possibly other techniques.
		\end{Remark}
	
	\begin{Remark} The proposed NN-based approach performs its derivations in real domain to determine the weights and bias through minimizing loss functions. However, a problem might arise that the complex derivations exist if and only if the loss functions satisfy the Cauchy-Riemann equations. In the complex domain, the functions that satisfy these equations are called holomorphic functions; otherwise, they are called non-holomorphic functions \cite{hirose2013complex}. This condition for complex domain introduces challenges for directly employing the proposed NN-based approach to learn to optimize in multiple antenna wireless communication systems. For example, in holographic multiple-input multiple-output (MIMO) surfaces and reconfigurable intelligent surfaces aided future wireless networks, the NNs need to deal with different parameters in complex domains. \end{Remark}

 Our aim is to utilize the high computational efficiency of the NN in its testing stage to design a time and computational efficient real-time power allocation scheme which can be applied to solve the power allocation problem with both perfect and imperfect CSI. As shown in Fig. 2, the proposed NN consists of three layers: input layer, multiple hidden layers and output layer. In particular, we choose $|\hat{h}_{s}|$, $|\hat{h}_{p}|$, $|\hat{h}_{e}|$, $|g_{s}|$, $|g_{e}|$, $\epsilon_{s}$, $\epsilon_{e}$ and $\epsilon_{p}$ as inputs and $P_{s}^{*}$ as output of the training data, respectively. Note that the perfect CSI scheme becomes a special case of imperfect CSI scheme by setting the inputs for the perfect CSI scheme as $|\hat{h}_{s}|=|h_{s}|$, $|\hat{h}_{p}|=|h_{p}|$, $|\hat{h}_{e}|=|h_{e}|$, and $\epsilon_{s}=\epsilon_{e}=\epsilon_{p}=0$. The mapping between the input and the output parameters can be expressed as 

\begin{align}\label{mapping_otimal}
P_{s}^{*}=f(|\hat{h}_{s}|, |\hat{h}_{p}|, |\hat{h}_{e}|, |g_{s}|, |g_{e}|, \epsilon_{s}, \epsilon_{e}, \epsilon_{p}).
\end{align}

We start from the input and then pass the input data through the NN and calculate the actual output straightforwardly, which is referred as feed-forward. Furthermore, the calculation flow follows the natural forward direction from the input layer to the hidden layers and finally to the output layer. This process can be expressed mathematically as

\begin{align}
\mathbf{z}^{(l+1)}=\mathbf{W}^{(l)}\mathbf{a}^{(l)}+\mathbf{b}^{(l)},
\end{align}
\begin{align}
\mathbf{a}^{(l+1)}=g(\mathbf{z}^{(l+1)}),
\end{align}
where $\mathbf{z}^{(l+1)}$ is the linear transformation of given inputs at the $(l+1)$-th layer, whereas $\mathbf{a}^{(l+1)}$ is the output activation value of the $(l+1)$-th layer. $g(\mathbf{z})$ denotes the activation function; in this work, we choose the rectified linear unit (ReLU) function as the activation function, which can be expressed as $g(x)=\max \{0,x\}$. $\mathbf{W}^{(l)}$ and $\mathbf{b}^{(l)}$ are the weight matrix and the bias vector for the $l$-th layer, respectively. Suppose there is an $N$-layer NN, the mapping between the inputs and the output parameters can be expressed as 
\begin{align}\label{mapping_neural}
y=f(\mathbf{S},\mathbf{W},\mathbf{b}),
\end{align}	
where $\mathbf{S}=[|\hat{h}_{s}|, |\hat{h}_{p}|, |\hat{h}_{e}|, |g_{s}|, |g_{e}|, \epsilon_{s}, \epsilon_{e}, \epsilon_{p}]$. Our goal is to determine the weights $\mathbf{W}=[\mathbf{W}^{(1)}, ...,$ $\mathbf{W}^{(N-1)}]$ and the bias $\mathbf{b}=[\mathbf{b}^{(1)},...,\mathbf{b}^{(N-1)}])$ such that both functions in (\ref{mapping_otimal}) and (\ref{mapping_neural}) yield a similar output for the same set of inputs. 

\begin{Proposition}\label{proposition:likelyhood_proof}
In order to have a similar outputs from both (\ref{mapping_otimal}) and (\ref{mapping_neural}), we should minimize the following normalized loss function: 

\begin{align}\label{lossfnc}
J(\mathbf{W},\mathbf{b})=\frac{1}{M}\sum_{m=1}^{M}(y_{m}-P_{s,m}^{*})^{2},
\end{align}

\noindent where $M$ is the number of training data sets. $y_{m}$ and $P_{s,m}^{*}$ are the $m$-th output of the NN and the optimal transmit power obtained by the conventional optimization approach, respectively. 
\end{Proposition}

\begin{IEEEproof}
	Please refer to Appendix \ref{proof_of_proposition}.
\end{IEEEproof}
~\\

We iteratively use the back-propagation based gradient descent algorithm to update the weights matrices $\mathbf{W}$ and the bias vectors $\mathbf{b}$. 
\begin{Proposition}\label{proposition:bp_proof}
	Based on the back-propagation and the gradient descent algorithm, the weight matrix and the bias vector for the $l$-th layer $\mathbf{W}^{(l)}$ and $\mathbf{b}^{(l)}$ can be updated respectively by 
	\begin{align}
	&\mathbf{W}^{(l)}=\mathbf{W}^{(l)}-\frac{\alpha}{M}\sum_{m=1}^{M}[\bm{\delta}_{m}^{(l+1)}(\mathbf{a}_{m}^{(l)})^{T}], \label{gdw}\\
	&\mathbf{b}^{(l)}=\mathbf{b}^{(l)}-\frac{\alpha}{M}\sum_{m=1}^{M}\bm{\delta}_{m}^{(l+1)},\label{bias}
	\end{align}
	where $\alpha$ is the learning rate and $\bm{\delta}_{m}^{(l+1)}$ is defined as $\bm{\delta}_{m}^{(l+1)}=\frac{\partial J(\mathbf{W},\mathbf{b})}{\partial \mathbf{z}_{m}^{(l+1)}}$. 
\end{Proposition}
\vspace{0.5 em}
\begin{IEEEproof}
	Please refer to Appendix \ref{proof_of_propositio1}.
\end{IEEEproof}
~\\

In an NN, over-fitting is the result of a model that is very closely to or precisely aligned to a specific set of data \cite{leinweber2007stupid}, which occurs when the model learns the training data set along with noises \cite{chicco2017ten}. Over-fitting leads the model not to be able to fit additional data or reliably predict future observations \cite{leinweber2007stupid}. Regularization is an approach to reduce the well-known over-fitting problem of a machine learning model \cite{bengio2011deep, girosi1995regularization}. To overcome this over-fitting problem, the $L_{1}$ and $L_{2}$ regularizations are most widely utilized techniques in the literature \cite{wang2006regularized, hoerl2000ridge}. 

The regularization term is added to the loss function to reduce the sum of absolute values of the weights in the $L_{1}$ regularization method, where the loss function can be written as 

\begin{align}\label{L1_reg}
J(\mathbf{W},\mathbf{b})\!=\!\frac{1}{M}\sum_{m=1}^{M}(y_{m}-P_{s,m}^{*})^{2}+\frac{\lambda}{2M}\sum_{l=1}^{N-1}||\mathbf{W}^{(l)}||_{1},
\end{align}
where $\lambda$ is the regularization parameter. Following the similar derivation of Proposition 2, the weights for the $L_{1}$ regularization can be updated as

\begin{equation}
\label{L1:weights}
\mathbf{W}^{(l)} = -\frac{\alpha}{M}\sum_{m=1}^{M}[
	\bm{\delta}_{m}^{(l+1)}(\mathbf{a}_{m}^{(l)})^{T}
]-
\frac{\alpha\lambda}{M}.
\end{equation}
 The bias $\mathbf{b}^{(l)}$ can be updated by using the equation provided in (\ref{bias}).

In the $L_{2}$ regularization method, the sum of squares of the weights are reduced by adding the regularization term to the loss function, which can be mathematically expressed as
\begin{align}\label{L2_reg}
J(\mathbf{W},\mathbf{b})\!=\!\frac{1}{M}\sum_{m=1}^{M}(y_{m}-P_{s,m}^{*})^{2}+\frac{\lambda}{2M}\sum_{l=1}^{N-1}||\mathbf{W}^{(l)}||_{2}^{2},
\end{align}
where $\lambda$ is the regularization parameter. Following a derivation similar to that of Proposition 2, the weights for the $L_{2}$ regularization can be updated as
\begin{align}\label{L2_weights}
\mathbf{W}^{(l)}\!=\!(1-\frac{\alpha\lambda}{M})\mathbf{W}^{(l)}\!-\!\frac{\alpha}{M}\sum_{m=1}^{M}[\bm{\delta}_{m}^{(l+1)}(\mathbf{a}_{m}^{(l)})^{T}].
\end{align}
 The bias $\mathbf{b}^{(l)}$ for $L_{2}$ regularization can be updated by using the equation provided in (\ref{bias}).

The development of our proposed NN scheme can be divided into three steps: (1) Obtaining the training data set by solving the secrecy rate maximization problem through conventional optimization approach; (2) developing an NN-based algorithm to learn the relationship between the input and output parameters of this secure transmission system; (3) after completing the training process, evaluating the performance of the trained NN over the conventional optimization algorithm. The details of these steps are provided in Algorithm 2.	
  \begin{algorithm}
  	\fontsize{10pt}{12pt}\selectfont
	\caption{\!\!: The NN approach}\label{alg:NN}
		\hspace*{0.02in} {\bf Preparing process:} 
	\begin{algorithmic}[1]
		\State Obtain the training data set by utilizing the conventional approaches in Section III: The optimal transmit power $P_{s}^{*}$ for corresponding the channel coefficients $|\hat{h}_{s}|, |\hat{h}_{p}|, |\hat{h}_{e}|, |g_{s}|, |g_{e}|$ and channel error bounds $\epsilon_{s}, \epsilon_{e}, \epsilon_{p}$;
	\end{algorithmic}
	\hspace*{0.02in} {\bf Training process:} 
\begin{algorithmic}[1]
		\State Initialize the weights matrices $\mathbf{W}$, the bias vectors $\mathbf{b}$ and the learning rate $\alpha$;
		\State Divide the training data set into $I$ mini-batches, the size of each mini-batch is $M$;
		\State \textbf{For each batch}: Input the training set $\mathbf{S}=[\mathbf{S}_{1},\ldots,\mathbf{S}_{M}]$ and $\mathbf{y}=[y_{1},\ldots,y_{M}]$, where $\mathbf{S}_{m}=[|\hat{h}_{s,m}|, |\hat{h}_{p,m}|, |\hat{h}_{e,m}|, |g_{s,m}|, |g_{e,m}|, \epsilon_{s,m}, \epsilon_{e,m}, \epsilon_{p,m}]$;
		\State For NN without any regularization, update the weights matrices $\mathbf{W}$ and the bias vectors $\mathbf{b}$ by minimizing the loss function defined in (\ref{lossfnc}) using the back-propagation based gradient descent method provided  in (\ref{gdw}) and (\ref{bias}); 
		\State For NN with $L_{1}$ regularization, update the weights matrices $\mathbf{W}$ and the bias vectors $\mathbf{b}$ by utilizing the back-propagation based gradient descent method provided in (\ref{L1:weights}) and (\ref{bias}), which are based on minimizing the loss function defined in (\ref{L1_reg});
		\State For NN with $L_{2}$ regularization, update the weights matrices $\mathbf{W}$ and the bias vectors $\mathbf{b}$ by minimizing the loss function defined in (\ref{L2_reg}) using the back-propagation based gradient descent method provided in (\ref{L2_weights}) and (\ref{bias});
		\State \textbf{End for};
		\State Save the trained NN.
	\end{algorithmic}
\hspace*{0.02in} {\bf Testing process:} 
\begin{algorithmic}[1]
\State Generate the channel coefficients for the test data set $\mathbf{S}_{test}$:
\State Feed $\mathbf{S}_{test}$ as the input parameters and determine the output results based on the trained NN;
\end{algorithmic}
\end{algorithm}
\section{Simulation Results}
In this section, we present numerical results to demonstrate the superior performance of our proposed NN schemes. The data set is obtained by utilizing the conventional optimization scheme in Section III with $6 \times 10^{5}$ different random channel realizations. We split the data set into two subsets of data: $5 \times 10^{5}$ for training and $10^{5}$ for validation. In the training process, all the NN parameters are updated by utilizing mini-batch gradient descent algorithm based on the Adam optimizer \cite{kingma2014adam}, where the batch size is chosen to be ten. All the parameters in NN are initialized with by the Xavier initializer \cite{glorot2010understanding}. Furthermore, similar to \cite{zhou2018resource}, it is assumed that the NN has two hidden layers with one hundred neurons in each layer. The learning rate $\alpha$ is set to $10^{-4}$ and the regularization parameter $\lambda$ is assumed to be $5\times 10^{-4}$ \cite{7926641,domingos2012few,wang2006regularized}. The test data set is obtained by using 3000 channel realizations. The transmit power of PU-Tx is assumed to be 60 mW, whereas all the noise variances are set to be 0.001. The channels $\hat{h}_{s}, \hat{h}_{p}, \hat{h}_{e}, g_{s}$, and $g_{e}$ are all generated by $\hat{h}_{i}=\chi_{i}\sqrt{d_{i}^{-\alpha}},~i=s,e,p$ and $g_{j}=\chi_{j}\sqrt{c_{j}^{-\alpha}},~j=s,e$, where $\chi_{i}\sim\mathcal{CN}(0,1)$, $ \chi_{j}\sim\mathcal{CN}(0,1)$, $d_{i}$ is the distance between the SU-Tx and the $i$-th user and $c_{j}$ denotes the distance between the PU-Tx and the $i$-th user. The parameter $\alpha=1.7$ denotes the path loss exponent. The distances between the transmitters and corresponding receivers are assumed to be $d_{s}=10$ m, $d_{e}=20$ m, $d_{p}=10$ m, $c_{s}=20$ m and $c_{e}=20$ m, respectively. The simulated datasets for training and testing were generated using MATLAB scripts, and the performance of data generation is irrelevant to the results. For training and testing the model, we used a system with Intel Core i7-9700K processor, with eight cores, clocked at 3.9 GHz, 12 MB cache memory and 32 GB random access memory. The training was performed purely on CPUs (opposed to GPUs).

First, we show the mean square error obtained by NN schemes without regularization, with $L_{1}$ regularization and $L_{2}$ regularization versus the  number of training steps, respectively, in Figs. 3-5. For a better presentation, we take samples for every 100 points from the whole training steps. It is obvious that the mean square error decreases and approaches zero as the number of iterations increases. This is due to the fact that the weights $\mathbf{W}$ and the bias $\mathbf{b}$ of the NN are iteratively updated by using the mini-batch gradient descent algorithm. Furthermore, the mean square errors of the validation data for the three schemes are also provided, respectively, in Figs. 3-5. As seen in these figures, the mean square errors first decrease and then remain constant, which confirms that the training process does not over-fit the NN for all three cases. Over-fitting is a phenomenon where a machine learning model becomes overly sensitive to a given dataset, and hence, fails to generalize beyond the training data~\cite{goodfellow2016deep,domingos2012few,tetko1995neural}. Generally, a model can easily be tested for over-fitting using a validation dataset that the model has not been exposed to. An over-fitted model has a signature characteristic of performing well for few training steps, and showing degrading performance for larger training steps~\cite{goodfellow2016deep,domingos2012few,tetko1995neural}, while the training performance increases. In our cases, it can be seen that the validation performance approaches a stead-state with increasing training steps. This is a clear evidence that the  model is not over-fitted.
	\begin{figure}[t]
		\centering\includegraphics[width=\linewidth]{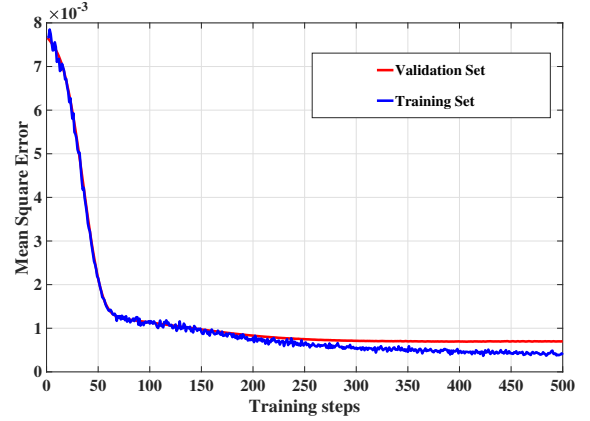}
		\caption{The mean square error between the power allocations obtained by the conventional approach and the NN scheme without regularization versus the number of training steps.}
	\end{figure}

	\begin{figure}[t]
		\centering\includegraphics[width=\linewidth]{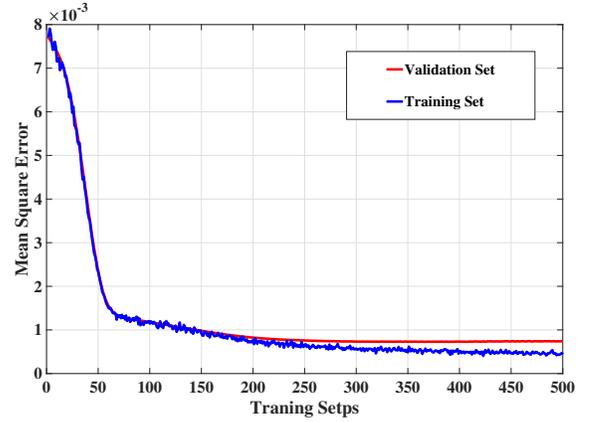}
		\caption{The mean square error between the power allocations obtained by the conventional approach and the NN scheme with $L_{1}$ regularization versus the number of training steps.}
	\end{figure}
 
Next, Fig. 6 presents the performance comparison in terms of optimal transmit power obtained by using the conventional optimization scheme and the proposed NN scheme (without regularization) versus the number of training steps. Similar to Figs. 3-5, the results of this figure are obtained by sampling every 100 points from the whole training steps. As seen in this figure, the output transmit power of the proposed NN scheme approaches the optimal transmit power obtained from the conventional scheme as the training steps increase. The reason is that the weights $\mathbf{W}$ and the bias $\mathbf{b}$ of the proposed NN are continuously updated in the training process to achieve minimum mean square error. Note that the output power of the proposed NN may be negative or larger than the available transmit power, since the training errors between the NN output power and the optimal power obtained by conventional optimization scheme cannot be completely eliminated. In order to incorporate the power constraints ($0\leq P_{s}\leq P_{t}$), we choose $P_{s}^{NN}=\min(\max(P_{s}^{OUT}, 0), P_{t})$ as the SU-Tx transmit power of our proposed NN scheme in the following simulation results.

	\begin{figure}[t]
		\centering\includegraphics[width=\linewidth]{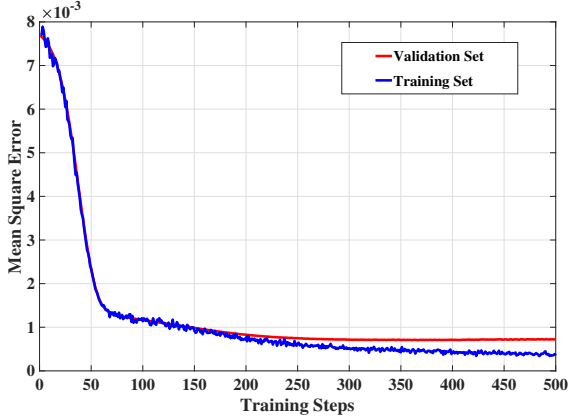}
		\caption{The mean square error between the power allocations obtained by the conventional approach and the NN scheme with $L_{2}$ regularization versus the number of training steps.}
	\end{figure}

	\begin{figure}[t!]
		\centering\includegraphics[width=\linewidth]{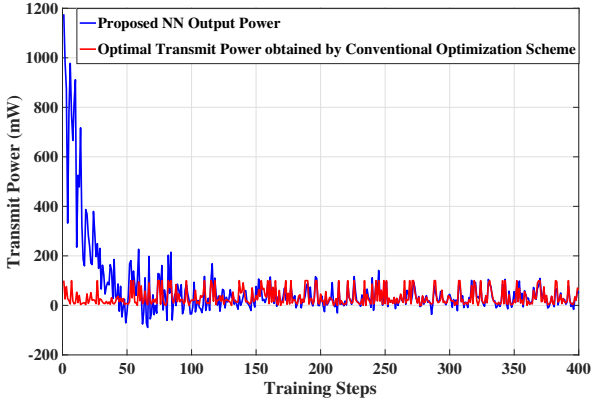}
		\caption{The performance comparison in terms of the optimal transmit power obtained by the conventional optimization approach and the proposed NN-based scheme without regularization versus the number of training steps.}
	\end{figure}

Next, Fig. 7 presents the achievable secrecy rates of the SU-Rx versus the interference leakage tolerance of the PU-Rx obtained by both conventional optimization and our proposed NN schemes with perfect CSI assumption. The maximum available transmit power of the SU-Tx is assumed to be 100 mW. It can be seen that the achievable secrecy rate increases with the interference leakage tolerance for all schemes. In addition, the three NN-based schemes can achieve a similar performance with the conventional optimization approach. Note that there is a performance gap between the conventional scheme and the three NN schemes, and this is due to the training errors between the output power and the desired optimal power.

	\begin{figure}[t]
		\centering\includegraphics[width=\linewidth]{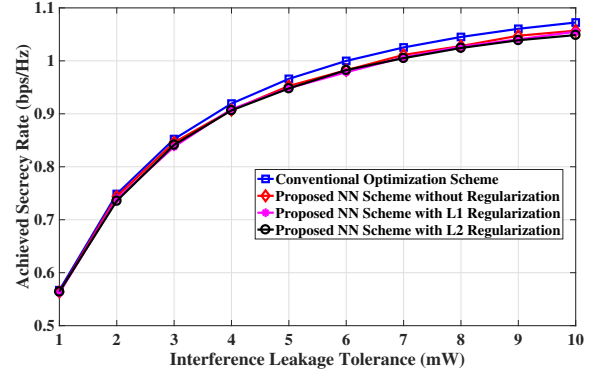}
		\caption{The achievable secrecy rates versus the interference tolerance of the PU-Rx obtained by the conventional optimization approach and the proposed NN framework under perfect CSI assumption.}
	\end{figure}

	\begin{figure}[t!]
		\centering\includegraphics[width=\linewidth]{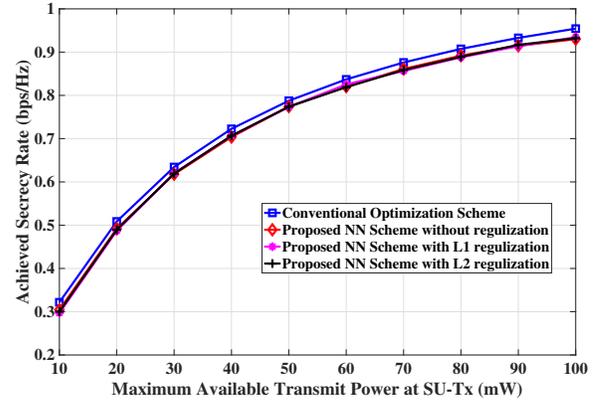}
		\caption{The achievable secrecy rates versus the maximum transmit power of the SU-Tx obtained by the conventional optimization approach and the proposed NN framework under perfect CSI assumption.}
	\end{figure}

Next, we evaluate the achievable secrecy rates versus the available transmit power with perfect CSI. Fig. 8 presents the achievable secrecy rates of SU-Rx of both conventional optimization and our proposed NN schemes. The interference leakage tolerance is set to 6 mW. It can be seen that the achievable secrecy rate increases as the transmit power enhances for all schemes. Similar to Fig. 7, our proposed NN schemes show a similar performance as the conventional optimization approach.

Next, Fig. 9 presents the achievable secrecy rates versus the interference leakage tolerance at the PU-Rx obtained by both conventional optimization and our proposed NN schemes under imperfect CSI assumption. The channel error bound is assumed to be $\epsilon_{s}=\epsilon_{e}=\epsilon_{p}=0.1$. The maximum available transmit power of SU-Tx is assumed to be 100 mW. As seen in Fig. 9, the achievable secrecy rate enhances as the interference leakage tolerance increases for all schemes. Furthermore, the three NN-based schemes show similar performances compared to that of the conventional optimization approach. 

	\begin{figure}[t]
		\centering\includegraphics[width=\linewidth]{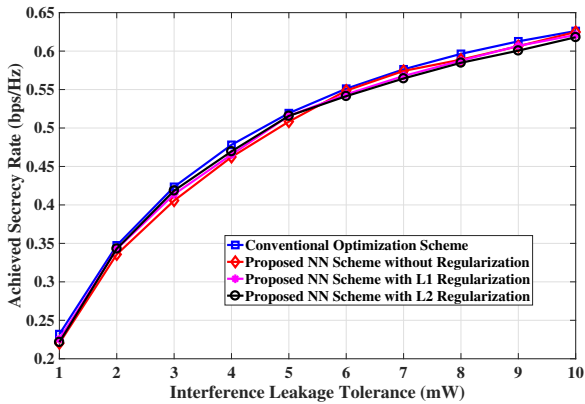}
		\caption{The achievable secrecy rates versus the interference tolerance of the PU-Rx obtained by the conventional optimization approach and the proposed NN framework under imperfect CSI assumption.}
	\end{figure}

	\begin{figure}[t]
		\centering\includegraphics[width=\linewidth]{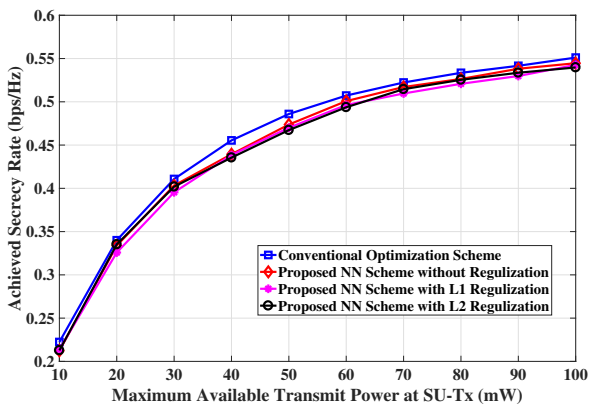}
		\caption{The achievable secrecy rates versus the maximum transmit power of the SU-Tx obtained by the conventional optimization approach and the proposed NN framework under imperfect CSI assumption.}
	\end{figure}

Next, we evaluate the achievable secrecy rates of  conventional optimization and the proposed NN schemes with different available transmit power at SU-Tx. Fig. 10 presents the achievable secrecy rates of SU-Rx for these schemes. The interference leakage tolerance is set to 6 mW. As seen in Fig. 10, the achievable secrecy rate enhances with the increase in the interference leakage tolerance. Similar to previous results, the proposed NN schemes provide a similar performance as the conventional convex optimization approach.

The achievable secrecy rates of conventional optimization and proposed NN schemes with different channel error bounds are provided in Fig. 11. The maximum available transmit power at SU-Tx is set to be 100 mW and the interference leakage tolerance at PU-Rx is assumed to be 6 mW. All the channel error bounds are assumed to be the same for each point, i.e., $\epsilon_{s}=\epsilon_{e}=\epsilon_{p}$. As seen in this figure, the achieved secrecy rate decreases as the channel error bound increases for all the schemes. Furthermore, as observed in the previous set of simulation results, the proposed NN schemes can achieve similar performances in comparison with the conventional convex optimization scheme.

	\begin{figure}[t]
		\centering\includegraphics[width=\linewidth]{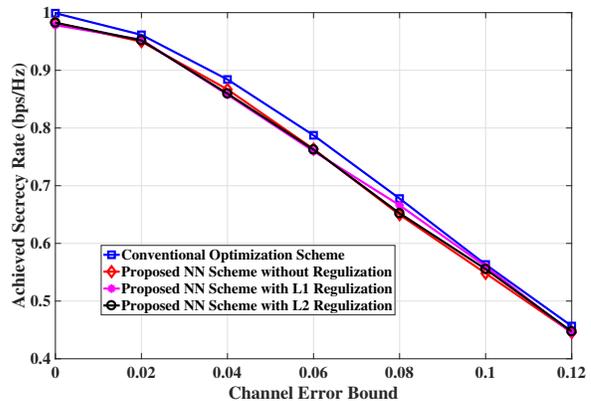}
		\caption{The achievable secrecy rates versus channel error bounds obtained by the conventional optimization approach and the proposed NN framework.}
	\end{figure}

	\begin{figure}[t]
		\centering\includegraphics[width=\linewidth]{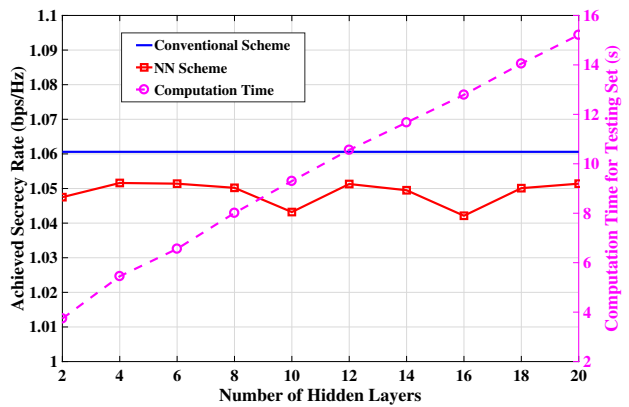}
		\caption{The achievable secrecy rates (left axis) and computation time for the training set (right axis) versus the number of hidden layers.}
	\end{figure}

In Fig. 12, we present the achieved secrecy rate (left axis) and computation time (right axis) versus the number of hidden layers for the NN scheme without any regularization. The maximum available transmit power at the SU-Tx and the interference leakage tolerance at the PU-Rx are assumed to be 100 mW and 9 mW, respectively. All channel error bounds are set to be 0 to represent the perfect CSI scenario. As shown in this figure, the difference of performance among different number of hidden layers are within a range of 1\%. However, the computation time for the testing set increases as the number of hidden layers increase. In other words, introducing more hidden layers cannot lead to much performance improvement, while it will yield more computational complexity to the NN.

	\begin{figure}[t]
		\centering\includegraphics[width=\linewidth]{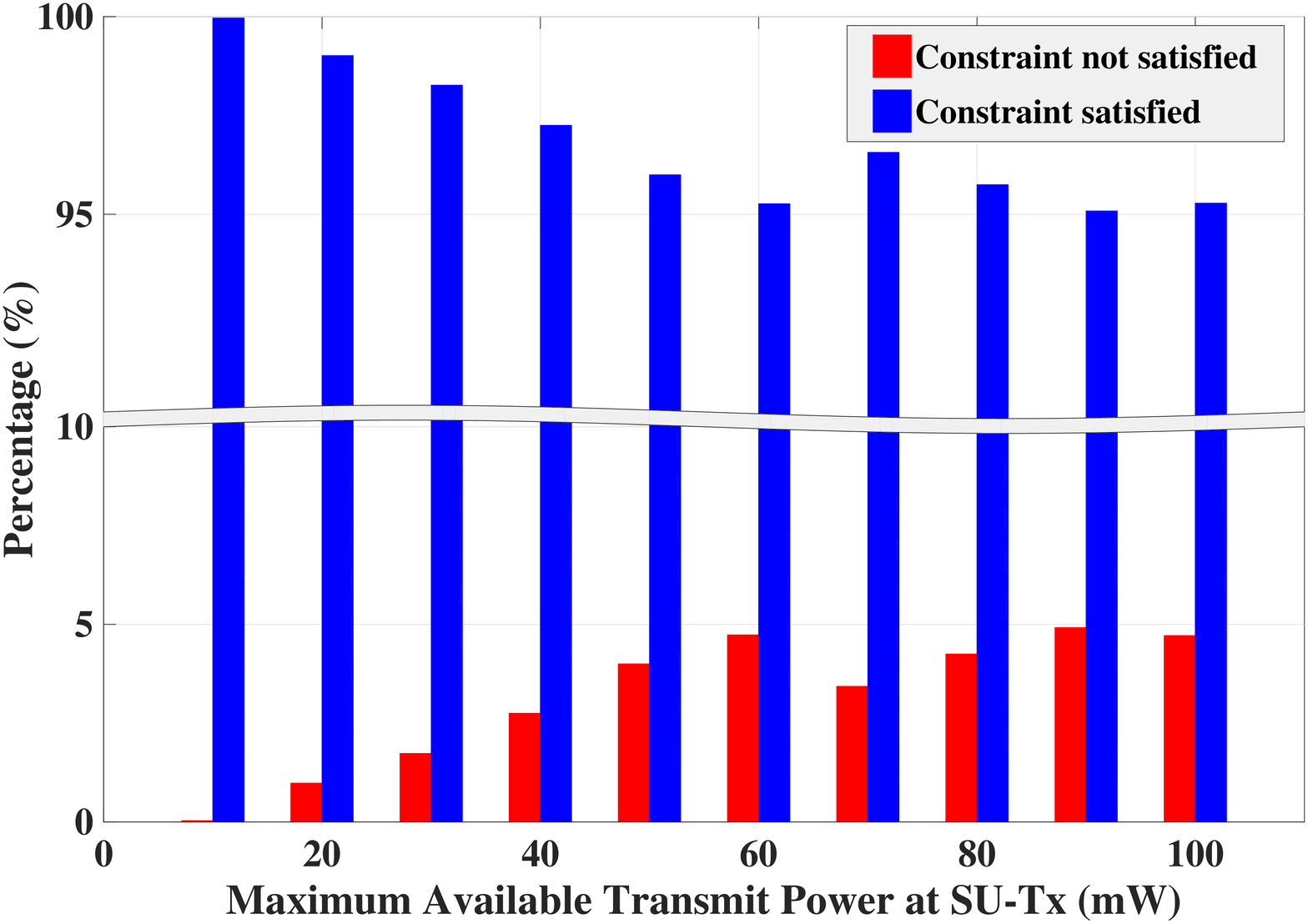}
		\caption{Distributions of the interference leakage satisfactions versus the maximum available transmit power at SU-Tx. }
	\end{figure}

	\begin{figure}[t]
		\centering\includegraphics[width=\linewidth]{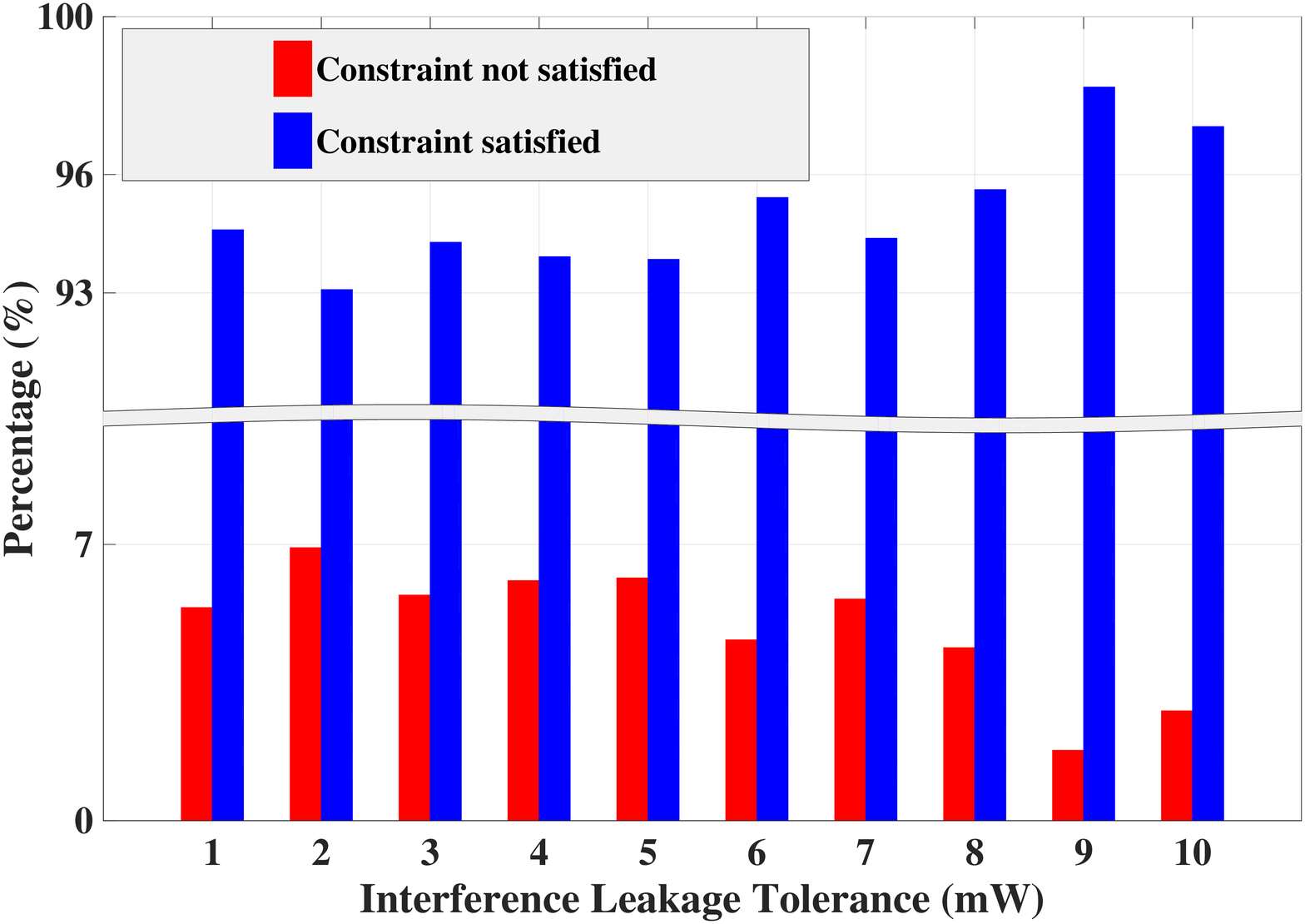}
		\caption{Distributions of the interference leakage satisfactions versus the interference leakages at RU-Rx. }
	\end{figure}

Next, we present the statistical results in Figs. 13 and 14 to evaluate the interference leakage tolerance satisfaction at the PU-Rx for the proposed NN scheme without regularization. These statistical results are calculated by combing the results of test data of both perfect and imperfect CSI scenarios. In Fig. 13, the interference leakage tolerance is set to 6 mW, while the maximum available transmit power at the SU-Tx is assumed to be 100 mW in Fig. 14. Fig. 13 provides the interference leakage tolerance satisfaction versus the maximum transmit power, whereas Fig. 14 presents it versus the interference leakage tolerances. As shown in these figures, more than 93\% of the test results can meet the interference leakage constraint at the PU-Rx.

\begin{table*}[ht!]
	\caption{The achieved secrecy rates of all schemes versus the interference leakage tolerances}
	\begin{center}	Perfect CSI\\
		\vspace{0.3 em}
		\begin{tabular}{|c|c|c|c|c|c|c|} 
			\hline 
			Interference leakage &NN scheme without& NN scheme with& NN scheme with & Conventional scheme &Minimum ratio \\
			tolerance & regularization & $L_{1}$ regularization &$L_{2}$ regularization &(bps/Hz)&(\%)\\
			(mW)&(bps/Hz)&(bps/Hz)&(bps/Hz)&&\\
			\hline
			1 &0.5636 &0.5618 &0.5642 &0.5679 &98.93\\
			\hline
			2&0.7438 &0.7393 &0.7355 &0.7486 &98.25\\
			\hline
			3 &0.846 &0.8377 &0.8414 &0.8523 &98.29\\
			\hline
		\end{tabular}
	\\~
	
	Imperfect CSI\\
	\vspace{0.3 em}
	\begin{tabular}{|c|c|c|c|c|c|c|} 
		\hline 
		Interference leakage &NN scheme without& NN scheme with& NN scheme with & Conventional scheme &Minimum ratio \\
		tolerance& regularization & $L_{1}$ regularization &$L_{2}$ regularization &(bps/Hz)&(\%)\\
		(mW)&(bps/Hz)&(bps/Hz)&(bps/Hz)&&\\
		\hline
		1 &0.2202 &0.2256 &0.2216 &0.2320 &94.91\\
		\hline
		2 &0.3357 &0.3427 &0.3431 &0.3474 &96.63\\
		\hline
		3 &0.4056 &0.4138 &0.4186 &0.4235 &95.77\\
		\hline
	\end{tabular}
	\end{center}
\end{table*}
\begin{table*}[ht!]
	\caption{The required computational time for all schemes versus the interference leakage tolerances}
	\begin{center}Perfect CSI\\
		\vspace{0.3 em}
		\begin{tabular}{|c|c|c|c|c|c|c|c|} 
			\hline
			Interference leakage &NN scheme without& NN scheme with& NN scheme with & Conventional scheme &Maximum ratio \\
			tolerance& regularization & $L_{1}$ regularization &$L_{2}$ regularization &(s)&(\%)\\
			(mW)&(s)&(s)&(s)&&\\
			\hline  
			1 &3.59 &4.40  & 4.38  &558.71 &0.79\\
			\hline 
			2 &3.69 & 4.58  &4.47 &584.68 &0.78\\
			\hline 
			3 &3.77 &4.45  &4.46 &573.38 &0.78\\
			\hline 
		\end{tabular}
	\\~
	
	Imperfect CSI\\
	\vspace{0.3 em}
	\begin{tabular}{|c|c|c|c|c|c|c|c|} 
		\hline
		Interference leakage &NN scheme without& NN scheme with& NN scheme with & Conventional scheme &Maximum ratio \\
		tolerance& regularization & $L_{1}$ regularization &$L_{2}$ regularization &(s)&(\%)\\
		(mW)&(s)&(s)&(s)&&\\
		\hline  
		1 &3.72 &4.67  & 4.22  &651.18 &0.65\\
		\hline 
		2 &3.71 & 4.79  &4.39 &639.32 &0.75\\
		\hline 
		3 &3.62 &4.74  &4.29 &647.28 &0.73\\
		\hline 
	\end{tabular}
	\end{center}
\end{table*}
\begin{table*}[ht!]
	\caption{The achieved secrecy rates of all schemes versus the maximum transmit powers}
	\begin{center}	Perfect CSI\\
	\vspace{0.3 em}
	\begin{tabular}{|c|c|c|c|c|c|c|} 
		\hline 
		Maximum available &NN scheme without& NN scheme with& NN scheme with & Conventional scheme &Minimum ratio \\
		transmit power& regularization & $L_{1}$ regularization &$L_{2}$ regularization &(bps/Hz)&(\%)\\
		(mW)&(bps/Hz)&(bps/Hz)&(bps/Hz)&&\\
		\hline
		10 &0.5636 &0.5618 &0.5642 &0.5679 &98.93\\
		\hline
		20 &0.7438 &0.7393 &0.7355 &0.7486 &98.25\\
		\hline
		30 &0.846 &0.8377 &0.8414 &0.8523 &98.29\\
		\hline
	\end{tabular}
\\~

	Imperfect CSI\\
	\vspace{0.3 em}
	\begin{tabular}{|c|c|c|c|c|c|c|} 
		\hline 
		Maximum available &NN scheme without& NN scheme with& NN scheme with & Conventional scheme &Minimum ratio \\
		transmit power& regularization & $L_{1}$ regularization &$L_{2}$ regularization &(bps/Hz)&(\%)\\
		(mW)&(bps/Hz)&(bps/Hz)&(bps/Hz)&&\\
		\hline
		10 &0.2123 &0.2138 &0.2127 &0.2224 &95.46\\
		\hline
		20 &0.3338 &0.3258 &0.3351 &0.3400 &95.82\\
		\hline
		30 &0.4033 &0.3954 &0.4019 &0.4108 &96.25\\
		\hline
	\end{tabular}
\end{center}
\end{table*}
\begin{table*}[ht!]
	\caption{The required computational time for all schemes versus the maximum transmit powers}
	\begin{center}Perfect CSI\\
		\vspace{0.3 em}
		\begin{tabular}{|c|c|c|c|c|c|c|} 
			\hline 
			Maximum available &NN scheme without& NN scheme with& NN scheme with & Conventional scheme &Minimum ratio \\
			transmit power& regularization & $L_{1}$ regularization &$L_{2}$ regularization &(s)&(\%)\\
			(mW)&(s)&(s&(s)&&\\
			\hline  
			10 &4.21 &5.03  &5.34  & 578.06 &0.92\\
			\hline 
			20 &3.90 &4.65  &4.77 &574.38 &0.83\\
			\hline 
			30 &3.63 &4.39 &4.58 &566.87 &0.81\\
			\hline 
		\end{tabular}
	\\~
	
	Imperfect CSI\\
	\vspace{0.3 em}
	\begin{tabular}{|c|c|c|c|c|c|c|} 
		\hline 
		Maximum available &NN scheme without& NN scheme with& NN scheme with & Conventional scheme &Minimum ratio \\
		transmit power& regularization & $L_{1}$ regularization &$L_{2}$ regularization &(s)&(\%)\\
		(mW)&(s)&(s)&(s)&&\\
		\hline  
		10 &3.87 &4.62 &4.76 & 658.36 &0.72\\
		\hline 
		20 &4.07 &4.52 &4.44 &654.87 &0.69\\
		\hline 
		30 &3.93 &4.82 &4.68 &661.43 &0.73\\
		\hline 
	\end{tabular}
	\end{center}
\end{table*}

Table I provides the secrecy rate performance of all schemes with perfect and imperfect CSI versus different interference leakage tolerances, similar to the results depicted in Figs. 7 and 9. Table II shows the comparison of the required computation time of the four schemes versus the interference leakage tolerances. Similarly, Table III presents the achieved the secrecy rate performance of all schemes versus the available transmit power. Table IV provides the required computation time of all schemes versus the maximum transmit power. To draw a performance comparison of the achieved secrecy rate, we employ the minimum ratio, which is calculated by dividing the minimum achieved secrecy rate among the three NN schemes by that of the conventional scheme. Similarly, for the comparison of the computational time, we use the maximum ratio, which is obtained by dividing the maximum computation of the three NN schemes by that of the conventional scheme. Note that the testing process for all schemes is performed on the same computer. For the results provided in these tables, the achievable secrecy rates are obtained by averaging results over the test data with 3000 channel realizations, while the computational time is the total computation time of 3000 channel realizations. From these results, we can conclude that the proposed NN schemes achieve at least 94\% of the optimal performance of the conventional scheme, while significantly reducing the required computation time. In particular, the proposed NN-based schemes require less than 1\% of the time needed by the conventional optimization scheme. This is due to the fact that the conventional optimization based solutions of the perfect CSI assumption are obtained through an iterative approach and sub-gradient algorithms, while the conventional scheme for the imperfect CSI assumption requires sub-gradient algorithms. These conventional optimization algorithms for both perfect and imperfect CSI scenarios are more complexity, which require a higher computation time. In the NN-based schemes, once the weights and bias are determined, it is should be able to compute the solution with a reasonable complexity within a short time compared to that of the conventional approach.
 \section{Discussions}
	Despite offering a number of benefits, the proposed approach also has a number of  shortcomings. We discuss these below, and highlight a number of potential directions for future work:
	\begin{enumerate}
		\item NN is a supervised learning approach, and hence relies on labelled-data for training the NN. In our context, the the necessity for valid labels implies that the training data should also be reliable, which is required for guaranteeing a valid solution~\cite{goodfellow2016deep}. In addition, the training is an off-line process, which effectively limits the applicability of the proposed approach to dynamic wireless systems. As we have mentioned in \emph{Remark 1}, hybrid approaches can be considered for problems in dynamic systems, which might include optimization techniques, NN, on-line learning, reinforcement learning and possibly other techniques.
		\item As mentioned in \emph{Remark 2}, the proposed NN approach may not be able to learn to optimize for multiple-antenna wireless transmission scenarios. To extend this NN-based scheme to multiple-antenna systems, one can consider two  approaches. One is to is to separate both complex input and output parameters into real and imaginary parts~\cite{8891562}, and the NN can be trained in real domain. The other is to handle the complex parameters by employing the Wirtinger calculus to deal with non-holomorphic functions in complex domain~\cite{hirose2013complex}.
		\item Finally, due to the fact that training errors cannot be completely eliminated, the proposed NN cannot include constraints in the training process, as presented in Fig. 13 and Fig. 14. This introduces challenges for extending the proposed scheme to design problems that have numerous system constraints. Fortunately, a constrained training algorithm was developed in the literature \cite{8792179}, where the key idea is to employ the Lagrange dual formulation to accommodate the constraints \cite{boyd2004convex}. This is another potential direction of future research work.
	\end{enumerate}

\section{Conclusion}	
In this paper, we proposed an NN-based approach for the power allocation design to maximize the secrecy rate in a CR network under transmit power and interference leakage constraints. We showed that the developed NN algorithm has the capability to solve the power allocation problem with both perfect and imperfect CSI, whereas it requires to develop both robust and non-robust optimization frameworks in the conventional approach. First, the conventional optimization scheme for perfect CSI scenario was developed based on a one-dimensional search, while that for the imperfect assumption was developed based on the Charnes-Cooper transformation and the S-Procedure approach. Then, the NN-based schemes were proposed where a relationship between the input and output parameters is established by determining an
approximated function. The training set to determine the relationship between inputs and output was obtained through the conventional optimization approaches and the NN was trained to calculate the weights of the connections in the network. After training the NN, the performance was evaluated with a test set in terms of achieved secrecy rate and required computational time. We demonstrated that the proposed NN schemes can achieve more than 94\% of the secrecy rate performance with less than 1\% computation time and more than 93\% satisfaction of interference leakage constraints compared with those of the conventional approaches. Simulation results were provided to demonstrate the effectiveness of the proposed NN-based approach over the benchmark conventional optimization approaches. Finally, we have discussed some limitations of the proposed NN-based approach and a number of potential future directions of research.

\begin{appendices}	
	\section{Proof of Proposition \ref{proposition:S_procedure}}
		First, we consider the following Lemma:		
	\begin{Lemma} (\emph{S-Procedure}~\cite{boyd1994linear}) Define $f_{i}(\mathbf{x})$, $i=1,2$ such as
		\begin{align} f_{i}(\mathbf{x})=\mathbf{x}^{H}\mathbf{A}_{i}\mathbf{x}+2Re\{\mathbf{b}_{i}^{H}\mathbf{x}\}+c_{i},
		\end{align}
		in which $\mathbf{x}\in\mathcal{R}^{n}$, $\mathbf{A}_{i}\in\mathcal{S}^{n}$, $\mathbf{b}_{i}\in\mathcal{R}^{n}$ and $c_{i}\in\mathcal{R}$. The implication $f_{1}(\mathbf{x})\leq 0 \rightarrow f_{2}(\mathbf{x})\leq 0$ holds if and only if there exists a $\vartheta\geq 0$ such that
		\begin{align}
		\vartheta\left[
		\begin{matrix}
		\mathbf{A}_{1} & \mathbf{b}_{1}\\
		\mathbf{b}_{1}^{H} & c_{1}\\
		\end{matrix}
		\right]
		-
		\left[
		\begin{matrix}
		\mathbf{A}_{2} & \mathbf{b}_{2}\\
		\mathbf{b}_{2}^{H} & c_{2}\\
		\end{matrix}
		\right]		 \succeq \mathbf{0}.
		\end{align}
	\end{Lemma}
	
	We first rewrite the constraint in (\ref{robust_C1}) as 
	\begin{align}
	&|e_{s}|^{2}-\epsilon_{s}^{2}\leq 0, \nonumber\\
	&\tau-t-\frac{\overline{P}_{s}|\hat{h}_{s}|^{2}+2Re\{\overline{P}_{s}\hat{h}_{s}e_{e}\}}+\overline{P}_{s}|e_{s}|^{2}{P_{p}|g_{s}|^{2}+\sigma_{s}^{2}}\leq 0.
	\end{align}
	Then, by applying Lemma 1, this constraint can be reformulated with a slack variable $\lambda_{1}$ as
	\begin{align}
	&\left[\
	\begin{matrix}
	\lambda_{1}+\frac{\overline{P}_{s}}{P_{p}|g_{s}|^{2}+\sigma_{s}^{2}} & \frac{\overline{P}_{s}\hat{h}_{s}}{P_{p}|g_{s}|^{2}+\sigma_{s}^{2}}\\
	\frac{\overline{P}_{s}\hat{h}_{s}}{P_{p}|g_{s}|^{2}+\sigma_{s}^{2}} & \frac{\overline{P}_{s}|\hat{h}_{s}|^{2}}{P_{p}|g_{s}|^{2}+\sigma_{s}^{2}}+t-\tau-\lambda_{1}\epsilon_{s}^{2}
	\end{matrix}
	\right]\succeq \mathbf{0},\nonumber\\
	&\lambda_{1}\geq 0.
	\end{align}
	Similarly, (\ref{robust_C2}) and  (\ref{robust_C3}) also can be derived as
	\begin{align}
	&|e_{e}|^{2}-\epsilon_{e}^{2}\leq 0, \nonumber\\
	&t-1+\frac{\overline{P}_{s}|\hat{h}_{e}|^{2}+2Re\{\overline{P}_{s}\hat{h}_{e}e_{e}\}+\overline{P}_{s}|e_{e}|^{2}}{P_{p}|g_{e}|^{2}+\sigma_{e}^{2}}\leq 0,
	\end{align}
	and 
	\begin{align}
	&|e_{e}|^{2}-\epsilon_{e}^{2}\leq 0, \nonumber\\
	&\overline{P}_{s}|\hat{h}_{p}|^{2}+2Re\{\overline{P}_{s}\hat{h}_{p}e_{p}\}+\overline{P}_{s}|e_{p}|^{2}-tq\leq 0,
	\end{align}
	respectively. Then, by adopting Lemma 1, these constraints can be reformulated, respectively as
	\begin{align}
	&\left[\
	\begin{matrix}
	\lambda_{2}-\frac{\overline{P}_{s}}{P_{p}|g_{e}|^{2}+\sigma_{s}^{2}} & -\frac{\overline{P}_{s}\hat{h}_{e}}{P_{p}|g_{e}|^{2}+\sigma_{s}^{2}}\\
	-\frac{\overline{P}_{s}\hat{h}_{e}}{P_{p}|g_{e}|^{2}+\sigma_{s}^{2}} & 1-\frac{\overline{P}_{s}|\hat{h}_{s}|^{2}}{P_{p}|g_{e}|^{2}+\sigma_{e}^{2}}-t-\lambda_{2}\epsilon_{e}^{2}
	\end{matrix}
	\right]\succeq \mathbf{0},\nonumber\\
	&\lambda_{2}\geq 0,
	\end{align}
	and 
	\begin{align}
	&\left[\
	\begin{matrix}
	\lambda_{3}-\overline{P}_{s} & -\overline{P}_{s}\hat{h}_{p}\\
	-\overline{P}_{s}\hat{h}_{p} &tq- \overline{P}_{s}|\hat{h}_{p}|^{2}+\sigma_{s}^{2}-\lambda_{e}\epsilon_{p}^{2}
	\end{matrix}
	\right]\succeq \mathbf{0},\nonumber\\
	&\lambda_{3}\geq 0.
	\end{align}
     This completes the proof of Proposition 1.	\hfill $\blacksquare$

	\section{Proof of Proposition \ref{proposition:likelyhood_proof}}\label{proof_of_proposition}
	In order to achieve a similar performance through the functions provided in (\ref{mapping_otimal}) and (\ref{mapping_neural}), we should maximize the following likelihood function:
	\begin{align}
	L(\mathbf{W},\mathbf{b})&=\prod_{m=1}^{M} p_{m}(y|x;\mathbf{W},\mathbf{b})\nonumber\\
	&=\prod_{m=1}^{M} \exp\bigg(-\frac{(f_{m}(\mathbf{S},\mathbf{W},\mathbf{b})-P_{s,m}^{*})^{2}}{2\sigma^{2}}\bigg).
	\end{align}
	By utilizing the monotonicity of the logarithmic function, the logarithmic likelihood function can be expressed as
	\begin{align}
	 &\log L(\mathbf{W},\mathbf{b})=\log \prod_{m=1}^{M} \exp\bigg(-\frac{(f_{m}(\mathbf{S},\mathbf{W},\mathbf{b})\!-\!P_{s,m}^{*})^{2}}{2\sigma^{2}}\bigg)\nonumber\\
	 &\!=\!M\!\log \bigg(\frac{1}{\sqrt{2\pi}\sigma}\bigg)\!-\!\frac{1}{2\sigma^{2}}\!\sum_{m=1}^{M}(f_{m}(\mathbf{S},\mathbf{W},\mathbf{b})\!-\!P_{s,m}^{*})^{2}.
	\end{align}
	Since $M\log \frac{1}{\sqrt{2\pi}\sigma}$ and $\frac{1}{2\sigma^{2}}$ are constants, maximizing the likelihood function is equivalent to minimizing the following loss function:
	\begin{align}
	J(\mathbf{W},\mathbf{b})=\sum_{m=1}^{M}(y_{m}-P_{s,m}^{*})^{2}.
	\end{align}
    Furthermore, this loss function can be normalized without loss of generality as follows:
		\begin{align}
	J(\mathbf{W},\mathbf{b})=\frac{1}{M}\sum_{m=1}^{M}(y_{m}-P_{s,m}^{*})^{2},
	\end{align}
	which completes the proof of Proposition 2.	 \hfill $\blacksquare$
	~\\
	\section{Proof of Proposition \ref{proposition:bp_proof}}\label{proof_of_propositio1}
	First, we provide the following basic chain rule:
	\begin{align}\label{chain}
	\frac{\partial h(g)}{\partial z}=\frac{\partial h}{\partial g}\frac{\partial g}{\partial z}.
	\end{align}
	From the feed-forward process, we have
	\begin{align}
	\mathbf{z}^{(l+1)}=\mathbf{W}^{(l)}\mathbf{a}^{(l)}+\mathbf{b}^{(l)},
	\end{align}
	\begin{align}
	\mathbf{a}^{(l+1)}=g(\mathbf{z}^{(l+1)}),
	\end{align}
	where $\mathbf{z}^{(l+1)}$ is the linear transformation of a given set of input parameters at the $(l+1)$-th layer, whereas $\mathbf{a}^{(l+1)}$ is the output activation value of the $(l+1)$-th layer. The function $g(\mathbf{z})$ represents the activation function. By assuming that $J(\mathbf{W},\mathbf{b})$ is the loss function of the NN, we can write the following equations based on the chain rule defined in (\ref{chain}):
	\begin{align}
	\frac{\partial J(\mathbf{W},\mathbf{b})}{\partial \mathbf{W}^{(l)}}&=\frac{\partial J(\mathbf{W},\mathbf{b})}{\partial \mathbf{z}^{(l+1)}}\frac{\partial \mathbf{z}^{(l+1)}}{\partial \mathbf{W}^{(l)}}\nonumber\\
	&=\frac{1}{M}\sum_{m=1}^{M}\bm{\delta}_{m}^{(l+1)}(\mathbf{a}_{m}^{(l)})^{T} ,
	\end{align}
	\begin{align}
	\frac{\partial J(\mathbf{W},\mathbf{b})}{\partial \mathbf{b}^{(l)}}&=\frac{\partial J(\mathbf{W},\mathbf{b})}{\partial \mathbf{z}^{(l+1)}}\frac{\partial \mathbf{z}^{(l+1)}}{\partial \mathbf{b}^{(l)}}=\frac{1}{M}\sum_{m=1}^{M}\bm{\delta}_{m}^{(l+1)}.
	\end{align}
	Since we can calculate $\mathbf{a}_{m}^{(l)}$ from the feed-forward process, then $\bm{\delta}_{m}^{(l)}$ can be derived as follows. Based on the chain rule, we have
	\begin{align}
	\bm{\delta}_{m}^{(l)}&=\frac{\partial J(\mathbf{W},\mathbf{b})}{\partial \mathbf{z}_{m}^{(l)}}=\frac{\partial J(\mathbf{W},\mathbf{b})}{\partial \mathbf{z}_{m}^{(l+1)}}\frac{\partial \mathbf{z}_{m}^{(l+1)}}{\partial \mathbf{a}_{m}^{(l)}}\frac{\partial \mathbf{a}_{m}^{(l)}}{\partial \mathbf{z}_{m}^{(l)}}\nonumber\\
	&=[(\mathbf{W}^{(l)})^{T}\bm{\delta}_{m}^{(l+1)}]\cdot g'(\mathbf{z}_{m}^{(l)}).	
	\end{align}
	Starting from the output layer, we can calculate $\bm{\delta}^{(l)}$ back forward  layer-by-layer until the input layer. Finally, considering the gradient descent method, the weights matrix $\mathbf{W}^{(l)}$ and the bias vector $\mathbf{b}^{(l)}$ for the $l$-th layer  can be updated respectively as follows:
	\begin{align}
	&\mathbf{W}^{(l)}=\mathbf{W}^{(l)}-\frac{\alpha}{M}\sum_{m=1}^{M}[\bm{\delta}_{m}^{(l+1)}(\mathbf{a}_{m}^{(l)})^{T}],\\
	&\mathbf{b}^{(l)}=\mathbf{b}^{(l)}-\frac{\alpha}{M}\sum_{m=1}^{M}\bm{\delta}_{m}^{(l+1)},
	\end{align}
	which completes the proof of of Proposition 3. \hfill $\blacksquare$
\end{appendices}

	
	%
	\bibliographystyle{IEEEtran}
	\bibliography{referenceIEEE}

\begin{thebibliography}{10}
\providecommand{\url}[1]{#1}
\csname url@samestyle\endcsname
\providecommand{\newblock}{\relax}
\providecommand{\bibinfo}[2]{#2}
\providecommand{\BIBentrySTDinterwordspacing}{\spaceskip=0pt\relax}
\providecommand{\BIBentryALTinterwordstretchfactor}{4}
\providecommand{\BIBentryALTinterwordspacing}{\spaceskip=\fontdimen2\font plus
\BIBentryALTinterwordstretchfactor\fontdimen3\font minus
  \fontdimen4\font\relax}
\providecommand{\BIBforeignlanguage}[2]{{%
\expandafter\ifx\csname l@#1\endcsname\relax
\typeout{** WARNING: IEEEtran.bst: No hyphenation pattern has been}%
\typeout{** loaded for the language `#1'. Using the pattern for}%
\typeout{** the default language instead.}%
\else
\language=\csname l@#1\endcsname
\fi
#2}}
\providecommand{\BIBdecl}{\relax}
\BIBdecl

\bibitem{cumanan2014secrecy}
K.~Cumanan, Z.~Ding, B.~Sharif, G.~Y. Tian, and K.~K. Leung, ``Secrecy rate
  optimizations for a {M}{I}{M}{O} secrecy channel with a multiple-antenna
  eavesdropper,'' \emph{IEEE Trans. Veh. Technol.}, vol.~63, no.~4, pp.
  1678--1690, May. 2014.

\bibitem{chu2015robust}
Z.~Chu, K.~Cumanan, Z.~Ding, M.~Johnston, and S.~Le~Goff, ``Robust outage
  secrecy rate optimizations for a {M}{I}{M}{O} secrecy channel,'' \emph{IEEE
  Wireless Commun. Lett.}, vol.~4, no.~1, pp. 86--89, Feb. 2015.

\bibitem{chu2016secrecy}
Z.~Chu, H.~Xing, M.~Johnston, and S.~Le~Goff, ``Secrecy rate optimizations for
  a {M}{I}{S}{O} secrecy channel with multiple multi-antenna eavesdroppers,''
  \emph{IEEE Trans. Wireless Commun.}, vol.~15, no.~1, pp. 283--297, Jan. 2016.

\bibitem{zhang2017secure}
M.~Zhang, K.~Cumanan, and A.~G. Burr, ``Secure energy efficiency optimization
  for {MISO} cognitive radio network with energy harvesting,'' in \emph{Proc.
  IEEE WCSP}, Nanjing, Oct. 2017, pp. 1--6.

\bibitem{Shannon}
C.~Shannon, ``Communication theory of secrecy systems,'' \emph{Bell Syst. Tech.
  J.}, vol.~28, no.~4, pp. 656--715, Oct. 1949.

\bibitem{Wyner}
A.~D. Wyner, ``The wire-tap channel,'' \emph{Bell Syst. Tech. J}, vol.~54,
  no.~8, pp. 1355--1387, Jan. 1975.

\bibitem{csiszar}
I.~Csisz{\'a}r and J.~Korner, ``Broadcast channels with confidential
  messages,'' \emph{IEEE Trans. Inf. Theory.}, vol.~24, no.~3, pp. 339--348,
  May 1978.

\bibitem{jiang2016machine}
C.~Jiang, H.~Zhang, Y.~Ren, Z.~Han, K.-C. Chen, and L.~Hanzo, ``Machine
  learning paradigms for next-generation wireless networks,'' \emph{IEEE
  Wireless Commun.}, vol.~24, no.~2, pp. 98--105, Apr. 2016.

\bibitem{goodfellow2016deep}
I.~Goodfellow, Y.~Bengio, and A.~Courville, \emph{Deep learning}.\hskip 1em
  plus 0.5em minus 0.4em\relax MIT Press Cambridge, 2016, vol.~1.

\bibitem{andrieu2003introduction}
C.~Andrieu, N.~De~Freitas, A.~Doucet, and M.~I. Jordan, ``An introduction to
  {MCMC} for machine learning,'' \emph{Machine learning}, vol.~50, no. 1-2, pp.
  5--43, Jan. 2003.

\bibitem{sebastiani2002machine}
F.~Sebastiani, ``Machine learning in automated text categorization,'' \emph{ACM
  Computing Surveys}, vol.~34, no.~1, pp. 1--47, Mar. 2002.

\bibitem{bishop2006}
C.~M. Bishop, \emph{Pattern Recognition and Machine Learning}.\hskip 1em plus
  0.5em minus 0.4em\relax Springer, 2006.

\bibitem{james1995}
J.~W. Kalat, \emph{Biological Psychology}.\hskip 1em plus 0.5em minus
  0.4em\relax Nelson Education, 1995.

\bibitem{chen2017machine}
M.~Chen, U.~Challita, W.~Saad, C.~Yin, and M.~Debbah, ``Machine learning for
  wireless networks with artificial intelligence: A tutorial on neural
  networks,'' \emph{IEEE Commun. Surveys \& Tutorials}, vol.~21, no.~4, pp.
  3039--3071, Jul. 2019.

\bibitem{demuth2014neural}
H.~B. Demuth, M.~H. Beale, O.~De~Jess, and M.~T. Hagan, \emph{Neural Network
  Design}.\hskip 1em plus 0.5em minus 0.4em\relax Martin Hagan, 2014.

\bibitem{8847377}
T.~{Lin} and Y.~{Zhu}, ``Beamforming design for large-scale antenna arrays
  using deep learning,'' \emph{IEEE Wireless Commun. Lett.}, vol.~9, no.~1,
  Jan. 2020.

\bibitem{8618345}
H.~{Huang}, Y.~{Song}, J.~{Yang}, G.~{Gui}, and F.~{Adachi},
  ``Deep-learning-based millimeter-wave massive {MIMO} for hybrid precoding,''
  \emph{IEEE Trans. Veh. Tech.}, vol.~68, no.~3, pp. 3027--3032, Mar. 2019.

\bibitem{8782877}
F.~{Zhou}, G.~{Lu}, M.~{Wen}, Y.~{Liang}, Z.~{Chu}, and Y.~{Wang}, ``Dynamic
  spectrum management via machine learning: State of the art, taxonomy,
  challenges, and open research issues,'' \emph{IEEE Network}, vol.~33, no.~4,
  Jul. 2019.

\bibitem{8761962}
C.~{Huang}, G.~C. {Alexandropoulos}, A.~{Zappone}, C.~{Yuen}, and M.~{Debbah},
  ``Deep learning for {UL/DL} channel calibration in generic massive {MIMO}
  systems,'' in \emph{Proc. ICC}, Shanghai, May 2019, pp. 1--6.

\bibitem{zhang2018robust}
M.~Zhang, K.~Cumanan, L.~Ni, H.~Hu, A.~G. Burr, and Z.~Ding, ``Robust
  beamforming for {AN} aided {MISO SWIPT} system with unknown eavesdroppers and
  non-linear {EH} model,'' in \emph{Proc. IEEE GLOBECOM WORKSHOP}, Abu Dhabi,
  Dec. 2018, pp. 1--7.

\bibitem{cumanan2017physical}
K.~Cumanan, H.~Xing, P.~Xu, G.~Zheng, X.~Dai, A.~Nallanathan, Z.~Ding, and
  G.~K. Karagiannidis, ``Physical layer security jamming: Theoretical limits
  and practical designs in wireless networks,'' \emph{IEEE Access}, vol.~5, pp.
  3603--3611, Dec. 2016.

\bibitem{cumanan2016secrecy}
K.~Cumanan, Z.~Ding, M.~Xu, and H.~V. Poor, ``Secrecy rate optimization for
  secure multicast communications,'' \emph{IEEE J. Sel. Topics Signal
  Process.}, vol.~10, no.~8, pp. 1417--1432, Dec. 2016.

\bibitem{chu2016simultaneous}
Z.~Chu, Z.~Zhu, M.~Johnston, and S.~Y. Le~Goff, ``Simultaneous wireless
  information power transfer for {M}{I}{S}{O} secrecy channel,'' \emph{IEEE
  Trans. Veh. Technol.}, vol.~65, no.~9, pp. 6913--6925, Nov. 2016.

\bibitem{chu2017robust}
Z.~Chu, K.~Cumanan, M.~Xu, and Z.~Ding, ``Robust secrecy rate optimisations for
  multiuser multiple-input-single-output channel with device-to-device
  communications,'' \emph{IET Commun.}, vol.~9, no.~3, pp. 396--403, Feb. 2015.

\bibitem{8649584}
M.~{Zeng}, N.~{Nguyen}, O.~A. {Dobre}, and H.~V. {Poor}, ``Securing downlink
  massive {MIMO-NOMA} networks with artificial noise,'' \emph{IEEE J. Sel.
  Signal Process.}, vol.~13, no.~3, pp. 685--699, Feb. 2019.

\bibitem{sun2017learning}
H.~Sun, X.~Chen, Q.~Shi, M.~Hong, X.~Fu, and N.~D. Sidiropoulos, ``Learning to
  optimize: Training deep neural networks for wireless resource management,''
  in \emph{Proc. IEEE SPAWC}, Sapporo, Jul. 2017, pp. 1--6.

\bibitem{zhang2019deep}
C.~Zhang, P.~Patras, and H.~Haddadi, ``Deep learning in mobile and wireless
  networking: A survey,'' \emph{IEEE Commun. Surveys \& Tutorials}, vol.~21,
  no.~3, pp. 2224-- 2287, Mar. 2019.

\bibitem{domingos2012few}
P.~M. Domingos, ``A few useful things to know about machine learning.''
  \emph{ACM Commun.}, vol.~55, no.~10, pp. 78--87, Spring 2012.

\bibitem{lecun2015deep}
Y.~LeCun, Y.~Bengio, and G.~Hinton, ``Deep learning,'' \emph{Nature}, vol. 521,
  no. 7553, pp. 436--444, May 2015.

\bibitem{ye2018power}
H.~Ye, G.~Y. Li, and B.-H. Juang, ``Power of deep learning for channel
  estimation and signal detection in {OFDM} systems,'' \emph{IEEE Wireless
  Commun. Lett.}, vol.~7, no.~1, pp. 114--117, Feb. 2018.

\bibitem{huang2019indoor}
C.~Huang, G.~C. Alexandropoulos, C.~Yuen, and M.~Debbah, ``Indoor signal
  focusing with deep learning designed reconfigurable intelligent surfaces,''
  in \emph{Proc. IEEE SPAWC}.\hskip 1em plus 0.5em minus 0.4em\relax IEEE,
  Cannes, Jul. 2019, pp. 1--5.

\bibitem{huang2020reconfigurable}
C.~Huang, R.~Mo, and C.~Yuen, ``Reconfigurable intelligent surface assisted
  multiuser {MISO} systems exploiting deep reinforcement learning,'' \emph{IEEE
  J. Sel. Commun.}, vol.~38, no.~8, pp. 1839 -- 1850, Aug. 2020.

\bibitem{zhou2018resource}
F.~Zhou, X.~Zhang, R.~Q. Hu, A.~Papathanassiou, and W.~Meng, ``Resource
  allocation based on deep neural networks for cognitive radio networks,'' in
  \emph{Proc. IEEE ICCC}, Beijing, Feb. 2018, pp. 40--45.

\bibitem{ye2018deep}
H.~Ye and G.~Y. Li, ``Deep reinforcement learning for resource allocation in
  {V2V} communications,'' in \emph{Proc. IEEE ICC}, Kansas City, May 2018, pp.
  1--6.

\bibitem{sun2018learning}
H.~Sun, X.~Chen, Q.~Shi, M.~Hong, X.~Fu, and N.~D. Sidiropoulos, ``Learning to
  optimize: Training deep neural networks for interference management,''
  \emph{IEEE Trans. Signal Process.}, vol.~66, no.~20, pp. 5438--5453, Oct.
  2018.

\bibitem{huang2019deep}
C.~Huang, G.~C. Alexandropoulos, A.~Zappone, C.~Yuen, and M.~Debbah, ``Deep
  learning for {UL/DL} channel calibration in generic massive {MIMO} systems,''
  in \emph{Proc. IEEE ICC}, Shanghai, May 2019, pp. 1--6.

\bibitem{sun2020machine}
Y.~Sun, J.~Liu, J.~Wang, Y.~Cao, and N.~Kato, ``When machine learning meets
  privacy in 6g: A survey,'' \emph{IEEE Communications Surveys \& Tutorials},
  Early Access 2020.

\bibitem{al2020survey}
M.~A. Al-Garadi, A.~Mohamed, A.~Al-Ali, X.~Du, I.~Ali, and M.~Guizani, ``A
  survey of machine and deep learning methods for {I}nternet of things
  ({I}o{T}) security,'' \emph{IEEE Communications Surveys \& Tutorials},
  Thirdquarter 2020.

\bibitem{o2017introduction}
T.~O’Shea and J.~Hoydis, ``An introduction to deep learning for the physical
  layer,'' \emph{IEEE Trans. Cognitive Commun. \& Network.}, vol.~3, no.~4, pp.
  563--575, Dec. 2017.

\bibitem{zhang2016secrecy}
M.~Zhang, K.~Cumanan, and A.~G. Burr, ``Secrecy rate maximization for
  {M}{I}{S}{O} multicasting {S}{W}{I}{P}{T} system with power splitting
  scheme,'' in \emph{Proc. IEEE SPAWC}, Edinburgh, Jul. 2016, pp. 1--5.

\bibitem{cumanan2017secure}
K.~Cumanan, G.~C. Alexandropoulos, Z.~Ding, and G.~K. Karagiannidis, ``Secure
  communications with cooperative jamming: Optimal power allocation and secrecy
  outage analysis,'' \emph{IEEE Trans. Veh. Technol.}, vol.~66, no.~8, pp.
  7495--7505, Aug. 2017.

\bibitem{charnes1962programming}
A.~Charnes and W.~W. Cooper, ``Programming with linear fractional
  functionals,'' \emph{Naval Res. Logist. Quart.}, vol.~9, no. 3-4, pp.
  181--186, 1962.

\bibitem{boyd1994linear}
S.~P. Boyd, L.~El~Ghaoui, E.~Feron, and V.~Balakrishnan, \emph{Linear {M}atrix
  {I}nequalities in {S}ystem and {C}ontrol {T}heory}.\hskip 1em plus 0.5em
  minus 0.4em\relax SIAM, 1994, vol.~15.

\bibitem{1199945}
S.~{Shahbazpanahi}, A.~B. {Gershman}, {Zhi-Quan Luo}, and {Kon Max Wong},
  ``Robust adaptive beamforming using worst-case sinr optimization: a new
  diagonal loading-type solution for general-rank signal models,'' in
  \emph{Proc. IEEE ICASSP}, vol.~5, Hong Kong, Apr. 2003, pp. V--333.

\bibitem{geraci2012secrecy}
G.~Geraci, M.~Egan, J.~Yuan, A.~Razi, and I.~B. Collings, ``Secrecy sum-rates
  for multi-user {MIMO} regularized channel inversion precoding,'' \emph{IEEE
  Trans. Commun.}, vol.~60, no.~11, pp. 3472--3482, Nov. 2012.

\bibitem{mukherjee2012detecting}
A.~Mukherjee and A.~L. Swindlehurst, ``Detecting passive eavesdroppers in the
  {MIMO} wiretap channel,'' in \emph{Proc. IEEE ICASSP}, Kyoto, Mar. 2012, pp.
  2809--2812.

\bibitem{1468482}
Y.~Guo and B.~C. Levy, ``Worst-case {MSE} precoder design for imperfectly known
  {MIMO} communications channels,'' \emph{IEEE Trans. Signal Process.},
  vol.~53, no.~8, pp. 2918--2930, Aug. 2005.

\bibitem{boyd2004convex}
S.~Boyd and L.~Vandenberghe, \emph{Convex {O}ptimization}.\hskip 1em plus 0.5em
  minus 0.4em\relax Cambridge University Press, 2004.

\bibitem{hornik1989multilayer}
K.~Hornik, M.~Stinchcombe, H.~White \emph{et~al.}, ``Multilayer feedforward
  networks are universal approximators.'' \emph{Neural networks}, vol.~2,
  no.~5, pp. 359--366, 1989.

\bibitem{reed2015neural}
S.~Reed and N.~De~Freitas, ``Neural programmer-interpreters,'' \emph{arXiv
  preprint arXiv:1511.06279}, 2015.

\bibitem{8626195}
J.~{Luo}, J.~{Tang}, D.~K.~C. {So}, G.~{Chen}, K.~{Cumanan}, and J.~A.
  {Chambers}, ``A deep learning-based approach to power minimization in
  multi-carrier {NOMA} with {SWIPT},'' \emph{IEEE Access}, vol.~7, pp.
  17\,450--17\,460, Jan. 2019.

\bibitem{jiang2017machine}
C.~Jiang, H.~Zhang, Y.~Ren, Z.~Han, K.-C. Chen, and L.~Hanzo, ``Machine
  learning paradigms for next-generation wireless networks,'' \emph{IEEE
  Wireless Commun.}, vol.~24, no.~2, pp. 98--105, Apr. 2017.

\bibitem{hirose2013complex}
A.~Hirose, \emph{Complex-valued {N}eural {N}etworks: Advances and
  {A}pplications}.\hskip 1em plus 0.5em minus 0.4em\relax John Wiley \& Sons,
  2013, vol.~18.

\bibitem{leinweber2007stupid}
D.~J. Leinweber, ``Stupid data miner tricks: overfitting the s\&p 500,''
  \emph{The Journal of Investing}, vol.~16, no.~1, pp. 15--22, Spring 2007.

\bibitem{chicco2017ten}
D.~Chicco, ``Ten quick tips for machine learning in computational biology,''
  \emph{BioData Mining}, vol.~10, no.~1, p.~35, Dec. 2017.

\bibitem{bengio2011deep}
Y.~Bengio, F.~Bastien, A.~Bergeron, N.~Boulanger-Lewandowski, T.~Breuel,
  Y.~Chherawala, M.~Cisse, M.~C{\^o}t{\'e}, D.~Erhan, J.~Eustache, X.~Glorot,
  X.~Muller, S.~P. Lebeuf, R.~Pascanu, S.~Rifai, F.~Savard, and G.~Sicard,
  ``Deep learners benefit more from out-of-distribution examples,'' in
  \emph{Proc. JMLR AISTATS}, Fort Lauderdale, Apr. 2011, pp. 164--172.

\bibitem{girosi1995regularization}
F.~Girosi, M.~Jones, and T.~Poggio, ``Regularization theory and neural networks
  architectures,'' \emph{Neural Computation}, vol.~7, no.~2, pp. 219--269, Mar.
  1995.

\bibitem{wang2006regularized}
L.~Wang, M.~D. Gordon, and J.~Zhu, ``Regularized least absolute deviations
  regression and an efficient algorithm for parameter tuning,'' in \emph{Proc.
  IEEE ICDM}, Hong Kong, Dec. 2006, pp. 690--700.

\bibitem{hoerl2000ridge}
A.~E. Hoerl and R.~W. Kennard, ``Ridge regression: biased estimation for
  nonorthogonal problems,'' \emph{Technometrics}, vol.~42, no.~1, pp. 80--86,
  Feb. 2000.

\bibitem{kingma2014adam}
D.~P. Kingma and J.~Ba, ``Adam: A method for stochastic optimization,''
  \emph{arXiv preprint arXiv:1412.6980}, 2014.

\bibitem{glorot2010understanding}
X.~Glorot and Y.~Bengio, ``Understanding the difficulty of training deep
  feedforward neural networks,'' in \emph{Proc, ICAIS}, Sardinia, 2010, pp.
  249--256.

\bibitem{7926641}
L.~N. {Smith}, ``Cyclical learning rates for training neural networks,'' in
  \emph{Proc. IEEE WACV}, Santa Rosa, Mar. 2017, pp. 464--472.

\bibitem{tetko1995neural}
I.~V. Tetko, D.~J. Livingstone, and A.~I. Luik, ``Neural network studies. 1.
  comparison of overfitting and overtraining,'' \emph{J. Chem. Inf. Comput.
  Sci.}, vol.~35, no.~5, pp. 826--833, Jan. 1995.

\bibitem{8891562}
T.~{Li}, M.~R.~A. {Khandaker}, F.~{Tariq}, K.~{Wong}, and R.~T. {Khan},
  ``Learning the wireless {V2I} channels using deep neural networks,'' in
  \emph{Proc. IEEE VTC-Fall}, Honolulu, Sep. 2019, pp. 1--5.

\bibitem{8792179}
H.~{Lee}, S.~H. {Lee}, and T.~Q.~S. {Quek}, ``Deep learning for distributed
  optimization: Applications to wireless resource management,'' \emph{IEEE
  Journal on Selected Areas in Communications}, vol.~37, no.~10, pp.
  2251--2266, Oct. 2019.

\end{thebibliography}

\end{document}